\documentclass[12pt]{article}
\pdfoutput=1

\usepackage{cite}
\usepackage{booktabs}
\usepackage[english]{babel}
\usepackage{amsmath,amssymb,amsbsy,amstext, amsthm, simplewick}
\usepackage{hyperref}
\usepackage{graphicx}
\usepackage{amsfonts}
\usepackage{amssymb}
\usepackage[small]{caption}
\usepackage{upgreek}
\usepackage[svgnames,dvipsnames,x11names,table]{xcolor}
\usepackage{multirow}
\usepackage{geometry}
\usepackage[hang,flushmargin]{footmisc}
\usepackage{bm}
\usepackage{braket}
\usepackage{subcaption}
\usepackage{mathtools}
\usepackage{setspace}
\usepackage{cleveref}
\usepackage{comment}
\usepackage{scalerel}
\usepackage[normalem]{ulem}
\usepackage{slashed}
\usepackage{enumitem}
\usepackage{dsfont}
\usepackage{tikz}
\usepackage[mathscr]{euscript}
\usetikzlibrary{decorations.markings}
\usetikzlibrary{shapes.misc}

\makeatletter
\g@addto@macro\bfseries{\boldmath}
\makeatother

\hypersetup{
    colorlinks=true,
    linkcolor={red!50!black},
    citecolor={blue!50!black},
    urlcolor={blue!80!black}
}

\usepackage[most]{tcolorbox}
\usepackage{amsmath}

\newtcolorbox[auto counter]{labeledbox}[2][]{
    enhanced,
    colback=gray!8,
    colframe=gray!50,
    boxrule=0.4pt,
    arc=3mm,
    width=\textwidth,
    left=8pt, right=8pt, top=8pt, bottom=8pt,
    overlay={
        \node[anchor=north west, font=\small\itshape, text=gray!60!black]
            at ([xshift=3mm,yshift=-2mm]frame.north west) {#2};
    },
    #1
}

\newtcolorbox{summarybox}[1][Summary of Results]{
    enhanced,
    colback=gray!8,
    colframe=gray!50,
    boxrule=0.4pt,
    arc=3mm,
    width=\textwidth,
    left=8pt, right=8pt, bottom=8pt, top=8pt,
    title=#1,
    fonttitle=\bfseries,
    colbacktitle=gray!8,     
    coltitle=black,
    toptitle=4pt, bottomtitle=4pt,
    titlerule=0pt,           
}

\usepackage{colortbl}

\makeatletter
\newlength{\apb@width}
\newcommand{\autoparbox}[2][c]{\settowidth{\apb@width}{#2}\parbox[#1]{\apb@width}{#2}}

\makeatother

\definecolor{lightgray}{gray}{0.9}

\usepackage[framemethod=default]{mdframed}
\newmdenv[skipabove=7pt,
skipbelow=7pt,
rightline=false,
leftline=false,
topline=false,
bottomline=false,
backgroundcolor=gray!10,
linecolor=gray,
innerleftmargin=5pt,
innerrightmargin=5pt,
innertopmargin=5pt,
innerbottommargin=5pt,
leftmargin=0cm,
rightmargin=0cm,
linewidth=4pt]{eBox}

\usepackage[most]{tcolorbox}
\tcbset{colback=white, colframe=black,
        highlight math style= {enhanced, 
            colframe=red,colback=red!10!white,boxsep=0pt}
        }
\definecolor{light-gray}{gray}{0.95}

\crefname{table}{Table}{Tables}
\crefname{equation}{Eq.}{Eqs.}
\crefname{appendix}{App.}{Apps.}
\crefname{section}{Sec.}{Secs.}
\crefname{figure}{Fig.}{Figs.}

\numberwithin{equation}{section}

\def\beq{\begin{equation}}
\def\eeq{\end{equation}}

\def\bea{\begin{eqnarray}}
\def\eea{\end{eqnarray}}

\def\beq{\begin{equation}}
\def\eeq{\end{equation}}
\def\bea{\begin{eqnarray}}
\def\eea{\end{eqnarray}}

\def\L{{\cal L}}

\def\O{{\cal O}}
\def\hO{{\hat {\cal O}}}
\def\bO{{\bar {\cal O}}}
\def\bo{{\bar {o}}}

\def\a{{\vec a}}

\def\k{{\vec{\scaleto{k}{7pt}}}}

\def\p{{\vec p}}

\def\x{{\vec x}}

\def\X{{x}}
\def\y{{\vec y}}

\def\L{{\cal L}}
\DeclareRobustCommand{\SkipTocEntry}[4]{}

\definecolor{colorTC}{rgb}{.2,.7,.2}

\definecolor{amethyst}{rgb}{0.6, 0.4, 0.8}

\definecolor{blue3}{RGB}{31, 119, 180}
\definecolor{red3}{RGB}{	214, 39, 40}
\definecolor{orange3}{RGB}{255, 127, 14}
\definecolor{green3}{RGB}{44, 160, 44}

\begin{document}

\begin{titlepage}
\setcounter{page}{1} \baselineskip=15.5pt
\thispagestyle{empty}
$\quad$
\vskip 70 pt

\begin{center}
{\fontsize{20}{18} \bf For Whom Does Bell Hold?}
\end{center}

\vskip 20pt
\begin{center}
\noindent
{\fontsize{12}{18}\selectfont Daniel Green, Kshitij Gupta, and Qiya Zhang}
\end{center}

\begin{center}
\vskip 4pt
\textit{{\small Department of Physics, University of California at San Diego,  La Jolla, CA 92093, USA}}

\end{center}

\vspace{0.4cm}
 \begin{center}{\bf Abstract}
 \end{center}

\noindent Violations of Bell's inequalities offer a definitive signal of non-classical (quantum) behavior in local deterministic systems. Yet, in many physical settings where quantum mechanics is expected to play an important role, one cannot construct a Bell-type test using the available observables. Cosmology offers one concrete example, where cosmic structure may have originated from quantum vacuum fluctuations yet all observations are effectively classical.  
Nevertheless, recent work suggests that quantum and classical time evolution may still be differentiated by the pattern of statistical correlations of these observables. In this paper, we explore and generalize the use of correlations as Bell-type test of the nature of time evolution. We show that quantum vacuum fluctuations of closed systems undergoing Hamiltonian evolution produce unique correlations that are not mimicked by classical Hamiltonian systems. Correlations in the quantum vacuum are generally time-independent and are controlled by the energy gap between the ground and exited states. Classical systems exhibit apparent poles at physical frequencies that do not arise in the quantum vacuum. As one approaches the poles, the evolution becomes dominated by a resonant Hamiltonian giving rise to time dependent correlations that eventually decay through dephasing. Finally, we show that this behavior is distinct from other signals of quantum evolution, including applications to quantum optics, quantum walks, and quantum search.





\end{titlepage}
\setcounter{page}{2}

\restoregeometry

\begin{spacing}{1.2}
\newpage
\setcounter{tocdepth}{2}
\tableofcontents
\end{spacing}

\setstretch{1.1}
\newpage

\section{Introduction}

The laws of physics, based on all observations to date, require quantum mechanics to hold on all scales from sub-atomic particles to the entire universe. Yet, not all phenomena require quantum mechanics to understand the system, either qualitatively or quantitatively. Most of the macroscopic world exists in this situation; we believe quantum mechanics underlies the laws, but we do need to use it to explain most phenomena larger than atomic scales\footnote{In other words, Bell's inequalities hold for thee \cite{donne1624devotions}.}. However, many of the most important problems in physics today reside in macroscopic settings where quantum mechanics is essential, including in black holes~\cite{Faulkner:2022mlp}, cosmology~\cite{Flauger:2022hie}, quantum matter~\cite{McGreevy:2016myw}, and quantum computing~\cite{Catterall:2022wjq}. Understanding how quantum mechanics manifests itself in these settings is therefore of both theoretical and experimental importance.

For simple spin systems, Bell's inequalities~\cite{Bell:1964kc,Bell:1964fg} and generalizations thereof~\cite{Clauser:1969ny,Greenberger:1989tfe}, offer an unambiguous window into the quantum nature of physical reality. Concretely, the  well-established experimental violation of these inequalities~\cite{Freedman:1972zza,Aspect:1982fx,Pan:2000djb,Stevens:2015awv,Handsteiner:2016ulx} categorically excludes the possibility of local hidden variables~\cite{Bohm:1951xw,Bohm:1951xx}. These tests ultimately exploit the difference between classical and quantum probability, where the former is defined by a set of real numbers with a one-norm while the latter is defined by complex numbers with a two-norm~\cite{Hardy:2001jk,Aaronson_2013}. While this difference in the definitions of probabilities is transparent in the mathematical expression of the physical systems, they cannot be distinguished for observables that commute. For commuting observables, the system can be diagonalized and the two types of probability become equivalent. The essence of Bell tests is that they involve measurements of local systems in non-commuting bases and thus expose the different definitions of probability~\cite{Clauser:1969ny}.

Unfortunately, testing quantum mechanics in macroscopic physical systems is a much greater challenge~\cite{Srednicki:1993im,Vidal:2002zz,Lu:2018yxh,Lu:2019xwg}. In many cases, all the practical experimental observables commute. For example, while position and momentum don't commute for a single quantum particle, the non-zero commutator of the macroscopic position and momentum of an object are suppressed by at least Avogadro's number~\cite{Gell-Mann:1992wkv,Zurek:2003zz,Banks:2008bd}. Although this is often an indication that quantum effects are irrelevant~\cite{Garrison:2015lva,Radicevic:2016kpf,Khemani:2017nda,Cohen:2020php}, we may still wish to experimentally interrogate the quantum nature of the system. The quantum origin of structure in the Universe provides an extreme example, where the largest objects in the Universe, galaxies and galaxy clusters, are thought to owe their existence to quantum fluctuations during inflation~\cite{Bardeen:1983qw,Guth:1982ec,Achucarro:2022qrl}. The observables today are classical by any practical definition~\cite{Grishchuk:1990bj,Campo:2005sv,Nelson:2016kjm,Martin:2017zxs,dePutter:2019xxv} and we cannot rewind the Universe to experimentally probe the inflationary era directly. Instead, we are left to infer the role of quantum mechanics from the classical statistical correlations that survive~\cite{Maldacena:2015bha}. Moreover, alternatives, such as warm inflation~\cite{Berera:1995ie,Berera:1998px,Berghaus:2019whh}, demonstrate that effectively classical fluctuations could be the true origin of structure and demand that we find a experimental probe that can distinguish these possibilities. Although the non-Gaussian correlators can be explicitly calculated for a wide range of quantum~\cite{Linde:1996gt,Maldacena:2002vr,Alishahiha:2004eh,Chen:2006nt,Cheung:2007st} and classical~\cite{LopezNacir:2011kk,Bastero-Gil:2014raa,Mirbabayi:2022cbt,Mirbabayi:2024eml,Salcedo:2024smn,Broadberry:2025ggb,Salcedo:2026sdn} examples, and tested against data~\cite{WMAP:2012fli,Planck:2019kim}, this approach alone cannot match the model-independence of a Bell test.

One hope is that time-evolution itself could pay the role of the non-commuting measurement in a conventional Bell-type test~\cite{Maldacena:2015bha,Green:2020whw}. In the quantum description, if the time-evolution cannot be diagonalized in the basis of the observable, then it is possible, in principle, for the evolution to differentiate between quantum and classical systems. Unfortunately, to reach the level of a Bell-type test, any proposal should be agnostic of the specific Hamiltonian. 
The most ambitious goal for this style of cosmic Bell test is to distinguish local deterministic classical evolution from quantum evolution without specifying additional details about the models.

At face value, it might seem impossible to characterize the space of all probability distributions that are obtainable by classical and quantum time evolution. Without further assumptions, there are no obvious limitations to the accessible distributions in either case. Fortunately, quantum computing (complexity) gives us a model for how this could be possible. It is widely believed that quantum computations can yield outcomes, given fixed resources, than are impossible for their classical counter-parts~\cite{Montanaro:2015taw,Harrow:2017jkm,Dalzell:2023ywa}. This is not because the final result involves non-commuting measurement, but because the quantum gates (time-evolution) allow the system to evolve to its final distribution in ways that are not accessible to a classical computer. For example, seeing the prime factorization of generic large numbers on a computer screen is a purely classical measurement by the observer, but is believed to require a computer that is exploiting the laws of quantum mechanics~\cite{Shor:1994jg}. At the very least, this suggests that assumptions about the size of the system or the length of time over which the system evolves may be sufficient to produce such a test. 

The signals of the quantum universe (SQU) proposal is a concrete attempt to live up to this goal~\cite{Green:2020whw,Green:2022fwg}. The essential idea is that the uncertainty principle permits the violation of energy conservation for a finite amount of time, thus allowing particles to be created from the vacuum. This is only possible in quantum mechanics when particle number does not commute with the Hamiltonian. It is a remarkable fact about local quantum field theory (QFT), that this feature can be inferred from the statistics of the quantum fluctuations in a way that does not depend on the details of the Hamiltonian. Unfortunately, the SQU proposal is uniquely tied to the framework of perturbative QFT and the relationship between correlation functions and scattering amplitudes~\cite{Baumann:2022jpr}. This adds a significant layer of scaffolding that makes the comparison to quantum mechanics, or even the definition of probability, more difficult.

In this paper, our goal is to generalize SQU to a purely quantum mechanical test. We will first review the original proposal and how it can be understood as a test of the quantum nature of Hamiltonian evolution. We then provide a more general construction in terms of a list of observables in a given finite dimensional theory, either quantum mechanical or classical. We focus on the difference between quantum vacuum fluctuations and classical fluctuations in a closed system undergoing local Hamiltonian evolution.

\begin{figure}[!ht]
    \centering
\includegraphics[width=0.85\textwidth]{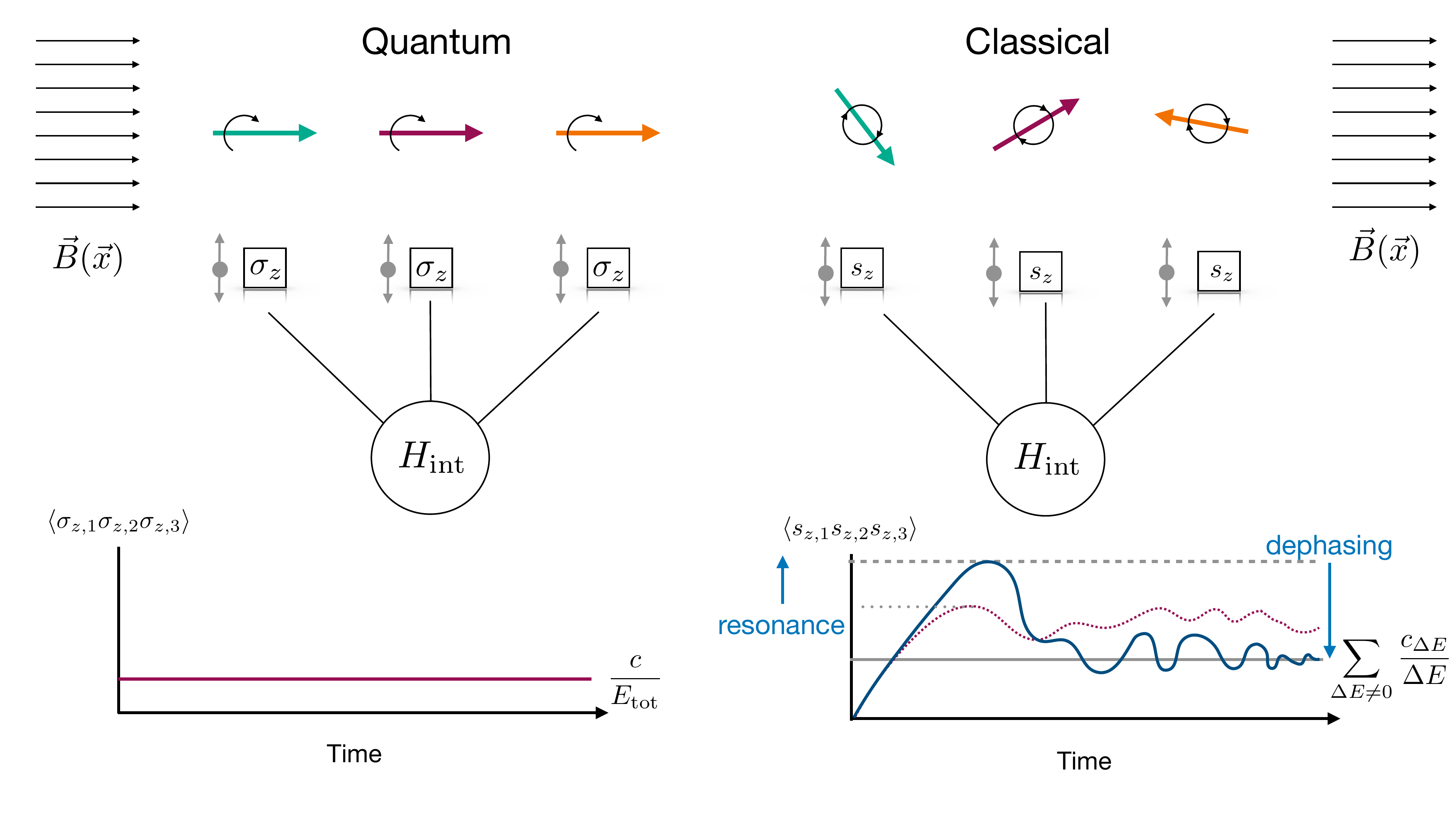}
    \caption{Illustration of the structural differences between quantum and classical theories. Both the quantum and classical theories have spins placed in the magnetic fields that cause spins to precess. The quantum theory (left) is placed in the ground state. The measurement of the spin in $z$-direction is random. When the spins are coupled via a Hamiltonian $H_{\rm int}$, the resulting correlations are time-independent with an amplitude set by the total energy $E_{\rm tot} = E_1+E_2+E_3$, where the energies $E_i$ are the gaps between the ground state and excited state of each spin. For the classical theory (right) to produce random spins in the $z$ direction, the spins are precessing in time with a frequency $E_i$ and a random phase. Coupling the spins generates a series of terms in the statistical correlations suppressed by all combinations of the frequency differences, $\Delta E = \pm E_1 \pm E_2 \pm E_3$, provided $\Delta E\neq 0$. Tuning the frequencies to a resonance, e.g.~$E_1 \to E_2 + E_3$, increases the correlations on short time scales, but causes the resonant pole to decay on long time scales through dephasing.  As $t\to \infty$, the classical correlations are still distinguished by the appearance of the off-resonant terms, $\Delta E \neq 0$, in the correlator. This example is studied explicitly in Section~\ref{subsec:spin}.
     }
    \label{fig:summary}
\end{figure}

The central difference between quantum and classical physics, in this setting, is that quantum systems are capable of exhibiting non-trivial statistical fluctuations of a set of observables while being in a unique vacuum state. This is possible precisely when the observables do not commute with the Hamiltonian. Moreover, since the vacuum state of the deformed theory remains stationary, one finds the correlators are time independent with predictable analytic properties.  In contrast, classical systems exhibiting the same two point statistics as the quantum theory, are necessarily excited into physical time-evolving states, with random amplitudes and phases. Any deformation of the classical theory allows these states to exchange energy, causing the correlations to grow on short time-scales. On long time scales, the correlations reach constant values, suppressed by the differences in frequencies, due to the dephasing of frequencies \cite{Kandrup:1998im}. If we tune the frequencies to a resonance, where frequencies are degenerate, a resonant Hamiltonian resolves the apparent pole, forcing the residue of the pole to vanish at late times. This binary difference in correlations is illustrated in Figure~\ref{fig:summary} for the example of spin-like observables. This qualitative difference is a reflection of the binary distinction between classical and quantum probability. We can continuously connect these two description working in a highly excited quantum state.

From this observation, we are drawn to central conclusions about SQU proposal. First, we find that the appearance of physical poles in the in-in correlation functions is the early time signal of resonant interactions. In fact, the apparent poles in the correlator reflect the failure of conventional perturbation theory and the need for degenerate perturbation theory. Second, using degenerate perturbation theory (i.e.~the resonant Hamiltonian), energy flows between the classical observables with frequencies that depend on the random classical variables. On long time scales, we average over these frequencies, which results in a vanishing correlation (i.e.~dephasing). In the limit of a large number of degrees of freedom, the decay is exponential in time, as anticipated in~\cite{Green:2020whw} from the decay of particles in quantum field theory.

\begin{summarybox}
\vskip -10pt 
Comparing the statistics of quantum vacuum fluctuations to classical statistical fluctuations in perturbed closed systems, we find the SQU proposal is generalized such that:
\begin{itemize}[itemsep=0pt, topsep=0pt]
\item Quantum vacuum correlations arise from the standard perturbative change to the vacuum. The ``total energy pole" arises from the factors of $(\rm{energy})^{-1}$  that appear in the mixing between the original ground state and excited states (\cref{eq:quantum_box} ).
\item Classical correlations involving a ``total energy pole" also exhibit correlations suppressed by the sum and differences of the frequencies (\cref{eq:offresonance}).
\item Deformations of the classical system give rise to a resonant Hamiltonian when combinations of frequencies are degenerate.  (\cref{eq:hprime}).
\item The resulting evolution generalizes and resolves the apparent physical poles that appear in SQU (\cref{eq:classical_box}).
\end{itemize}
\end{summarybox}

A natural question, given this construction, is how it relates to existing tests of quantum mechanics. We review several proposals that share qualitative similarities to look for sharp connections between SQU and other results. We discuss the relationship with the negativity of pseudo-probability distributions such as the Glauber $P$-representation and the Wigner distribution. Although both offer sharp contrasts between specific quantum and classical systems, we show neither offers a similar sharp test for commuting observables. A more natural comparison with our proposal is quantum-speedups, and we explore examples where proposed quantum speed-ups  might share the same physical origin as the SQU test. Although we do not find a general connection, our results suggest natural directions for future work at this intersection.

This paper is organized as follows: In Section~\ref{sec:review}, we review the SQU proposal and its relationship to Bell-tests. In Section~\ref{sec:spectra}, we demonstrate that quantum and classical theories differ in both their spectra and their correlations. In Section~\ref{sec:relation} we show that in limit of large degrees of freedom, we see exponential decay of classical correlations, as anticipated in the SQU proposal. In Section~\ref{sec:compare}, we compare this proposal with existing quantum tests. We conclude in Section~\ref{sec:conclusions}.

\section{Statistics and Evolution}\label{sec:review}

Non-commuting observables are the ultimate origin of quantum behavior, and Bell's inequalities expose the consequences of such observables is an unambiguous way. Bell tests require that we perform measurements that don't commute in systems where the outcomes are binary. Unfortunately, most macroscopic physical systems don't fit this requirement. Yet, calculations of quantum phenomena on any scale require non-commuting variables and thus contain the quantum nature of reality. The question is whether the differences in the way we calculate classical and quantum correlators leaves any meaningful differences in commuting observables. In this section, we will review the nature of quantum versus classical time evolution and its connection to statistical correlations.

\subsection{Bell Tests and Commutators}

Mathematically, there are two possible definitions of probability that define an internally self-consistent physical system~\cite{Hardy:2001jk,Aaronson_2013},
\bea
p_{\rm classical}(X) &= \sum_{i\in X} p_i \qquad \quad &p_i \in [0,1] \qquad \sum_{i} p_i = 1 \ , \\
p_{\rm quantum} (X) &= \left|\sum_{i\in X} \psi_i\right|^2 \qquad \quad &\psi_i \in {\mathbb C} \qquad \left|\sum_{i} \psi_i\right|^2 = 1 \ .
\eea
In other words, both probabilities are defined by a sum of events that lead to the same physical outcome, $X$, but the classical and quantum definitions are distinguished by whether the states define an $\ell^1$ or $\ell^2$ space respectively. 

Given this binary difference, it would seem like it should be straightforward to isolate the role of quantum probabilities in physical systems. However, we observe neither $p_i$ nor $\psi_i$, and therefore we must construct a physical observable that is sensitive to their differences. If we consider a set of mutually commuting observables $\O_k$ with $[\O_k,\O_j]=0$, the we can always diagonalize $\psi_i$ with respect to $\O_k$ so that there is a unique orthogonal state associated to every unique eigenvalue $\omega_{k,i}$. Therefore, the probability of observing the $i$-th eigenvalue of $\O_k$, $\omega_{k,i}$, is 
\beq
p_{\rm quantum} (\{\omega_{k,i}\} ) =  |\psi_{k,i}|^2 \to p_{k,i} \ .
\eeq
Because these states are all physically distinct (orthogonal), there are no interference terms and we can reproduced any observable defined in terms of $\O_k$ with a purely classical probability distribution $p_{k,i}$. Therefore, in order to isolate the role of quantum mechanics, we must also be sensitive to a non-zero commutator.

A violation of Bell's inequalities is possible precisely because they require a set of observables $a,a'$ and $b,b'$ where $[a,a'] \neq 0$ and $[b,b']\neq 0$. Specifically, if the eigenvalues for all these observables are required to be $a,a',b,b'=\pm 1$, then we can define an observable sensitive to the commutators~\cite{Clauser:1969ny}
\beq
\hat C = ab+a' b +a' b'- a b' \quad \to \quad \hat C^2 = 4 + [a,a'][b,b'] \ .
\eeq
We see that the statistical correlations of $a,a'$ and $b,b'$ are sensitive to their commutator and therefore a classical system must obey the CHSH inequality $|\hat C|\leq 2$. A quantum theory violates this inequality and generates $C^2 > 4$. The Bell and CHSH inequalities are also used to express the importance of entanglement as a probe of quantum mechanics.

There are two key challenges with using Bell-inequalities in many physical settings. The most transparent is that we must be able to measure two non-commuting observables by repeated measurement of a physically identical state. Complex physical system are difficult to engineer to be in an exact quantum state. Moreover, we do not always have access to non-commuting observables. The second difficulty is that Bell's tests assume we know the exact list of eigenvalues ahead of time and that they are identical for all realizations of the quantum or hidden variables description. In larger systems, it is often not the case that the distribution of eigenvalues is known ahead of time. For example, if $a,a',b,b' = \pm \lambda$ for some (possibility random) number $\lambda$, then the CHSH bound,
\beq
C^2 = 4 \lambda^2 + [a,a'][b,b'] \ ,
\eeq
is only useful if we can determine $\lambda$ independently. This problem becomes more severe as we move to continuous variables with unknown probability distribution functions.

\subsection{Time-Dependence and the Quantum Universe Test}

In many physical situations, our observables are limited to a single basis. For any given operator, $\hO$, we diagonalize the states in the eigen-basis so that
\beq
|\psi \rangle =\sum_i c_i |\psi_i \rangle \qquad \hO |\psi_i \rangle =\lambda_i  |\psi_i \rangle  \qquad \langle \psi_i | \psi_j \rangle = \delta_{ij} \ .
\eeq 
In this situation, the states are always orthogonal for distinct eigenvalues. For illustration, if we have a quantum mechanical particle that we can only study through measurements of position, $\hat x$, then any statistical quantity we measure, $f(\hat x)$, will produce the same result as a classical system where $p(x) = |\psi(x)|^2$.

However, in situations of this kind, we might expect that the space of probability distributions $p(x) = |\psi(x)|^2$ that can be achieved via quantum time evolution is not the same as those distributions that can be reached by purely classical evolution. If we work in the Heisenberg picture, the time evolution of an observable in a unitary theory is
\beq
\frac{d}{dt} {\cal O} = \frac{\partial}{\partial t} {\cal O} + \frac{i}{\hbar} [H, {\cal O}] \ .
\eeq
Here we have introduced $\hbar$ so that we can understand the classical limit. Certainly if $[H,
\O] =0$, then the time evolution will not contain a non-zero commutator and thus will be purely classical. In fact, in this limit, ${\cal O}$ will not meaningfully evolve at all and therefore the statistics of $\O$ reveal nothing about the time evolution, classical or quantum. The factor of $\hbar^{-1}$ is a reminder that one commutator survives the classical limit so that 
\beq
\lim_{\hbar \to 0} \frac{d}{dt} {\cal O} \to \frac{\partial}{\partial t} {\cal O} + \{ H , \O \} \qquad \ \{ H , \O \} = \partial_x H \partial_p \O - \partial_p H \partial_x \O \ .
\eeq
In other words, the classical equations of motion arise from a single commutator between $q$ and $p$, $[q,p] = -i \hbar$.

The time evolution of a quantum theory can, in principle, be distinguished from a classical theory when there are still non-commutating operators beyond the classical limit, i.e. 
\beq
[H, \O] = -i \hbar \O' \qquad [\O, \O'] \neq 0 \ . 
\eeq
In this case, the evolution equation for our observable $\O$ necessarily involves states that are not diagonal in the $\O$ basis. As a result, the evolution cannot be trivially described as a classical process. This basic observation suggests, in principle, that the time evolution of a quantum and classical system can be fundamentally different in a way that could allow for a novel test of the quantum nature of a given system.

There are many reasons to suspect such a test is possible. Quantum speedups in various computing problems indeed suggest that we could tell is someone replaced classical gates with quantum gates from the novel outcomes that become possible with a quantum computer~\cite{Shor:1994jg,Grover:1997fa,Montanaro:2015taw}. Unfortunately, in most physical systems of interest, the time evolution is not under our control at a microscopic level, and therefore we need tests that work in more generic settings\footnote{One can construct models where the Universe performs a Bell-test during the inflation and stores the outcome in some statistical quantity~\cite{Maldacena:2015bha}. While this formally achieves our goal, on some level, it requires the laws of nature to be highly contrived. }.

The Leggett-Garg tests \cite{Emary:2013wfl} serve as one example of time evolution providing a Bell like test, where, in the simplest form, one measures an unequal time correlation function of an operator ${\cal O}(t)$ as $C_{12} = \langle {\cal O}(t_1) {\cal O}(t_2) \rangle$ at three different times. If ${\cal O}$ can only have the binary output $\pm 1$, then, one can show that the classical inequality $C_{12} + C_{23} - C_{13} \leq 1$ is violated by the quantum theory. These tests also require the presence of a non-trivial Hamiltonian which does not commute with ${\cal O}$ in order to generate violations. While Legett-Garg tests don't require measurements of non-commuting variables, we still need to have knowledge of the eigenvalues of the measurements beforehand. Moreover, it can be shown that the Legett-Garg tests and Bell-type test both follow from the same core idea that non-commuting variables cannot have definite values at different spatial or temporal separations\cite{Markiewicz:2014qpw, Das:2014cud}. 

The signals of a quantum universe (SQU) test proposed in~\cite{Green:2020whw} is another concrete attempt to realize this goal. It's utility lies in being able to distinguish classical and quantum signatures based on statistical measurements at a fixed time. The key input is the understanding of how local quantum field theory and causality control the analytic structure of the correlations. The test was originally specific to inflationary backgrounds, but a slightly simpler description can be given in terms of a set of local particle detectors in flat space~\cite{Green:2022fwg}. Suppose we have a scalar field $\phi$, which in the limit of no interactions is given by
\beq\label{eq:phiQSU}
\phi(\x, t)=\int \frac{d^3 k}{(2 \pi)^3} e^{i \k \cdot \x} \frac{1}{\sqrt{2 k}}\left[a^\dagger_{\mkern-2mu-\vec{k}} e^{i k t}+a_{\k} e^{-i k t}\right]  \ .
\eeq
The quantum theory in a ground state $|0\rangle$, is defined the operators $a_\k$ and $a_\k^\dagger$ satisfying,  
\beq\label{eq:qstats}
\left[a_{\vec{k}}, a_{\overrightarrow{k^{\prime}}}^{\dagger}\right]=(2 \pi)^3 \delta\left(\vec{k}-\vec{k}^{\prime}\right) \quad a_{\vec{k}}|0\rangle=0 \ .
\eeq
For a classical theory, the amplitudes are statistical variables defined by
\beq\label{eq:cstats}
\left\langle a_{\vec{k}}^{\dagger} a_{\overrightarrow{k^{\prime}}}\right\rangle_c=\frac{1}{2}(2 \pi)^3 \delta\left(\vec{k}-\vec{k}^{\prime}\right)=\left\langle a_{\overrightarrow{k^{\prime}}} a_{\vec{k}}^{\dagger}\right\rangle_c .
\eeq
By construction, both the theories have the same equal time two-point statistics.

Now we introduce an interaction via the Hamiltonian,
\beq
H_{\rm int} = \frac{1}{3!} \mu \phi^3 \ .
\eeq
The quantum mechanical in-in correlator can be calculated perturbatively in the interaction picture as
\beq
\phi_I = e^{-i \int dt' H_{\rm int}} \phi  e^{i \int dt' H_{\rm int}}\ . 
\eeq
The resulting bispectrum is given by 
\bea
\langle \phi(\k_1) \phi(\k_2) \phi(\k_3) \rangle_Q &=&2 \operatorname{Im} \int_{-\infty}^t d t^{\prime} \frac{\mu}{8 k_1 k_2 k_3} e^{-i\left(k_1+k_2+k_3\right)\left(t-t^{\prime}\right)} \\
& =&-\frac{\mu}{4 k_1 k_2 k_3\left(k_1+k_2+k_3\right)} \ .
\eea
The key feature of this formula is the pole in the total energy $k_T = k_1+k_2+k_3$. The residue of the pole is given by the tree level scattering amplitude~\cite{Maldacena:2011nz,Raju:2012zr,Benincasa:2018ssx}, but the existence of the pole itself is controlled by kinematics. The presence of this pole reflects that this is the creation of 3 particles from the vacuum.

The classical bispectrum can be calculated using the same Hamiltonian evolution, but where the correlators follow from the statistics of the linearized field using the Gaussian two-point statistics in Equation~(\ref{eq:cstats}). At linear order in $\mu$, the field is given by 
\beq
\phi^{(2)}(\vec{k}, t)=\frac{\mu}{2} \int^t d t^{\prime} G\left(k ; t-t^{\prime}\right) \int \frac{d^3 p}{(2 \pi)^3} \phi^{(1)}\left(\vec{p}, t^{\prime}\right) \phi^{(1)}\left(\vec{k}-\vec{p}, t^{\prime}\right) \ .
\eeq
where $\phi^{(1)}$ is the linearized field and $G\left(k ; t-t^{\prime}\right)$ is the retarted Green's function. The resulting bispectrum is given by
\bea
\langle \phi(\k_1) \phi(\k_2) \phi(\k_3) \rangle_C &=&
\langle \phi^{(2)}(\k_1) \phi^{(1)}(\k_2) \phi^{(1)}(\k_3) \rangle_C + {\rm permutations} \\
&=& \frac{\mu}{16 k_1 k_2 k_3}\left(\frac{3}{k_T}+\sum_{i=1}^3 \frac{1}{k_T-2 k_i}\right) \ . \label{eq:biclass}
\eea
The noteworthy difference in the classical case is the poles at physical momenta where $k_1 = k_2 +k_3$ or permutations thereof. These poles reflect the appearance of $1 \to 2$ particle decay, and $2\to 1$ particle annihilation processes. The fluctuations of $\phi$ in the classical theory reflect real excitations and therefore all allowed processes occur. In the quantum theory, these processes are only prevented from occurring in vacuum because there are no particles in the initial state, by definition.

It is critical to this proposal that it is not merely the existence of a non-zero bispectrum, but its analytic structure, that defines the test. It is straightforward to construct examples where, given a fixed Hamiltonian, the bispectrum vanishes in the classical limit. Of course, our goal is not to test $\hbar \neq 0$ in a fixed theory with a given Hamiltonian, but as a statement about the space of all local theories. A compelling reason to focus on the analytic properties of the correlators is that they are robust to the details of the interactions~\cite{Baumann:2022jpr}.

The weakness of the SQU test is that it seems to rely heavily on the specific structure of quantum field theory and the types of observables that are possible for a field $\phi(\x,t)$ that is local and causal. The argument for the appearance of physical poles is intimately connected to the vanishing of commutators outside the light-cone, much like the need for anti-particles in a relativistic quantum field theory~\cite{Feynman1987Antiparticles}. While this is the relevant observable for cosmology, it does not share the ubiquity of a quantum test as Bell's inequalities. Our goal now is to understand which features of this proposal can be generalized outside of the context of perturbative quantum field theory.

Additionally, the appearance of poles at physical momenta suggests some kind of failure of our perturbative description. Such a pole would represent infinite signal to noise, meaning the classical and quantum cases are perfectly distinguishable with any experimental sensitivity. It was already noted in~\cite{Green:2020whw} that the physical poles should be interpreted as a decay process, and therefore we would expect the pole to be resolved by a finite width (imaginary self-energy) for each mode. However, without a general understanding of how these poles are resolved, one cannot directly determine the experimental sensitivity that would be needed to execute the SQU test.

\subsection{The Classical Limit}\label{subsec:classical}

As we have established, quantum and classical mechanics are not smoothly connected theories. The rules for probability in the two descriptions are unique and not related by a continuous parameter. Of course, we also experience the quantum universe as being effectively classical most of the time, and therefore the quantum description must become classical in a controlled way. We would like to show how this works in the context of the SQU test, following the approach of~\cite{Green:2022fwg}.

The key idea is that the SQU test only distinguishes the classical theory from the quantum ground state. The classical limit, as is well known, must therefor involve large occupation numbers. In fact, it has long been known that tree-level correlations are enhanced in excited states~\cite{Holman:2007na}. There are multiple ways to achieve this but in the interest of being most transparent, we will excite the vacuum directly with some number of modes
\beq
| \Omega\rangle \to |\Phi\rangle = \int \frac{d^3 k}{(2\pi)^3} \frac{1}{\sqrt{n_\k!}} (a_\k^\dagger)^{n_k} |0\rangle  \ ,
\eeq
where $n_\k$ are integers. This state is defined to be a particle number eigenstate for each of the modes:
\beq
\hat n_\k |\Phi\rangle =  \langle a_\k^\dagger a_\k |\Phi\rangle =  n_\k |\Phi \rangle ,
\eeq
where the factor of $n_\k$ comes from repeated use of the commutator to eliminate $a_\k$.

It will also be useful allow for the overall normalization of the field $\phi$ to be adjusted with a free parameter, $f_\phi$, so that 
\beq\label{eq:phinorm}
\phi(\x, t)=\int \frac{d^3 k}{(2 \pi)^3} e^{i \k \cdot \x} \frac{f_\phi}{\sqrt{2 k}}\left[a^\dagger_{\mkern-2mu-\vec{k}} e^{i k t}+a_{\k} e^{-i k t}\right],
\eeq
This is helpful for keeping the normalization of the two point statistics fixed, but can be absorbed into a change of units.

Now let us look at the statistics of $\phi(\k,t)$ in this state (in the quantum description). The power spectrum is given by 
\beq
\langle \phi(\k,t) \phi(\k',t) \rangle = \frac{1}{2k} f_\phi^2 (2 n_k +1) \ .
\eeq
Taking $f_\phi^2 (2 n_k +1) =1$ keeps the power spectrum fixed, assume $n_\k \to n$ is a constant for the observable modes.

The perturbative bipsectrum can, in principle, be calculated in the same way as the quantum theory:
\beq
\langle \phi(\k_1) \phi(\k_2) \phi(\k_3) \rangle=2 \operatorname{Im} \int_{-\infty}^t d t^{\prime} \frac{\mu f_\phi^6}{8 k_1 k_2 k_3} \prod_{i=1}^3\left( (n+1)e^{-i k_i\left(t-t^{\prime}\right)}+ ne^{i k_i\left(t-t^{\prime}\right)} \right)  \ .
\eeq
While evaluating this expression is messy, it is easy to see that all of the operator ordering associated with the quantum theory is associated with the factors of $n+1$ vs $n$. As a result, if we take the limit $n \to \infty$ holding $f_\phi^2 n$ and $f_\phi^2 \mu$ fixed, we get at leading order
\bea
\lim_{n\to \infty} \langle \phi(\k_1) \phi(\k_2) \phi(\k_3) \rangle_C = \frac{\mu f_\phi^6 n^2 }{16 k_1 k_2 k_3}\left(\frac{3}{k_T}+\sum_{i=1}^3 \frac{1}{k_T-2 k_i}\right) \ . \label{eq:biclasslim}  
\eea
The result is sensitive to how the time integrals are evaluated, as it is in the classical theory, but the observation that we recover the classical statistics in the limit of large occupation holds regardless. 

\section{Generalized Quantum Universe Tests}\label{sec:spectra}

In the previous section, we established the idea that Hamiltonian evolution could play the role of a Bell-test. We further showed how such an idea could be realized in (quantum) field theory via the SQU proposal. 

The SQU test depends on two basic assumptions about the origin of statistical correlations:
\begin{enumerate}
\item Two-point correlations of the ``observable" quantities of both the classical and quantum theories should agree.
\item Higher-point correlations can be generated by a local interaction Hamiltonian. 
\end{enumerate}
These two conditions should be understood as analogues of the assumptions of the original Bell tests. Condition 1 is the analogue of requiring $a,a',b,b' =\pm 1$, which establishes the two point statistics unambiguously. Condition 2 is the time-evolution equivalent of local realism, namely that the correlations we observe are explainable by local physics. 

Our goal in this section is to remove the scaffolding of cosmology and quantum field theory, and reproduce the analytic signatures of SQU in the case of ordinary quantum mechanics. At the same time, generalizing the proposal beyond perturbative QFT demands that we make sense of the poles that arise at physical energies. We will see that this signals the need for resumming the divergences into a time dependent term via the resonant Hamiltonian, hence helping us understand the structure of the correlator both on and off the resonance poles. 

To achieve our goal, we will formalize what we mean by the two SQU conditions in the context of general closed systems. We will then use this setup to diagnose the difference between a quantum and classical evolution. In order to gain intuition for our general construction, we will start with a simple example that illustrates the key physical ideas. 

\subsection{Warm-up Example}\label{subsec:spin}

We will consider a system with three independent spins $s_1, s_2, s_3$ in magnetic fields $\vec B_{i=1,2,3} = B_{i} \hat x$. This example is illustrated visually in Figure~\ref{fig:summary}. We will assume that we have sufficient control of the magnetic field to produce a resonance where $B_1 =B_2+B_3$.
\vskip 5pt
\noindent {\bf Quantum Evolution:} In the quantum theory, we will take these to be a spin-$1/2$ with a Hamiltonian 
\beq
H =  - \sum_i \vec B_i \cdot \vec \sigma_i \ ,
\eeq
where $\vec \sigma_i = (\sigma_i^x,\sigma_i^y, \sigma_i^z)$ are the Pauli matrices that act on the spin $i$. We will put the spins in the ground state
\beq
|\Psi \rangle  = \frac{1}{2\sqrt{2}} \prod_{i=1}^3 e^{-i B_i t}\begin{pmatrix}
1 \\
1
\end{pmatrix}_i \equiv | \rightarrow, \rightarrow, \rightarrow \rangle  \ .
\eeq
Now suppose we measure the spin in the $\hat z$ direction, $\sigma_z$, we will find for each of the spins
\beq
\langle \prod_i \hat z \cdot \sigma_i \rangle = 0 \qquad \langle (\hat z \cdot \sigma_i)^2 \rangle  = 1 \ .
\eeq
Now we turn on the interaction to generate a correlation
\beq
H_{\rm int} = g \prod_i \hat z\cdot \vec \sigma_{i} = g \sigma_1^z \sigma_2^z \sigma_3^z \ .
\eeq
We can use conventional perturbation theory to understand its impact on the state,
\beq
|\delta \Psi\rangle = \frac{\langle \leftarrow,\leftarrow,\leftarrow | g \sigma^z_{1}\sigma^z_{2} \sigma^z_{3} | \rightarrow, \rightarrow, \rightarrow \rangle }{2 (B_1+B_2 +B_3)} | \leftarrow,\leftarrow,\leftarrow \rangle \ .
\eeq
Here we see the sum over energies of the individual states, as excepted. This mixing with the original high energy state then generates the correlation
\beq \label{eq:SSSquant}
\langle \sigma_1^z \sigma_2^z \sigma_3^z \rangle  = \frac{g}{B_1 + B_2 + B_3} + {\cal O}(g^2) \ .
\eeq
Since the frequencies (energies) are given by $B_i$, we see the existence of the total energy pole for the quantum vacuum, as expected. The interaction doesn't do much to the system because the states are not degenerate. We slightly perturb the system in a way that correlates the spins, but the energy and state remain very close to the ground state of the unperturbed system due to the gap between the ground state and the fully excited state, $2(B_1+B_2+B_3)$. 
\vskip 5pt
\noindent {\bf Classical Evolution:} Now we compare the structure of quantum correlations to the classical correlators. We again have three spins, $\vec s_i$, precessing in fields $\vec B_{i=1,2,3} = \hat x B_{i=1,2,3}$. To parallel the quantum measurements, we will focus on the $z$-component, whose evolution is given by
\beq
\vec s \cdot \hat z \equiv S_i = I_i \cos\theta_i =  I_i \cos( B_i t + \varphi_i) \ ,
\eeq
where $I_i$ and $\theta_i$ are the action angle variables, where $\theta_i =B_i t + \varphi_i$  and $I_i =$ constant are the solutions to the equations of motion. To match the quantum statistics, we will take $I_i = 2$ and $\varphi_i$ is uniformly distributed between $[0,2\pi)$. We will keep $I_i$ because it defines an action variable in the Hamiltonian description. 

We can check the statistics of the $z$-components
\beq
\langle \prod_i S_i \rangle = 0 \qquad \langle S_i^2 \rangle  = 1 \ .
\eeq
Now, in detail, we might notice that $S_i$ is not simply $\pm 1$, but we can define an observable that only measures the sign of $S_i$  that would recover the same properties as the quantum case.

Again, we introduce the interaction,
\beq
H_0 = \sum_i B_i I_i \qquad H_{\rm int} = g S_1 S_2 S_3 \ .
\eeq
Note here that we simplified\footnote{The correct Hamiltonian for the complete classical system is the same as in the quantum system. However, determining the classical equations of motion requires keeping all the spin components, $\vec s_i$ and the so(3) algebra to recover the correct equations of motion.  We will instead work with a Hamiltonian for $S_i$ alone that reproduces the same equations of motion as the complete description.} our description to three pairs of action angle variables, $I_i$ and the azimuthal angles $\theta_i$.

Now, we want to calculate the same correlation function $\langle S_1 S_2 S_3 \rangle$ as we did for the quantum case. Perturbatively, we get that 
\begin{equation}
    \delta S_1 = - g \int dt\, G_1(t - t')  S_2 S_3 \qquad G_i(t - t') = \Theta(t -t ') \frac{\sin (B_i (t - t'))}{B_i} \ .
\end{equation}
with similar corresponding equations for $\delta S_2, \delta S_3$. Using the statistical correlation of the undeformed theory, 
\begin{equation}
    \langle S_i(t) S_i(t') \rangle = \frac{\langle I_i\rangle^2}{2} \cos (B_i (t - t')) \qquad I_i^2 = 1 \ ,
\end{equation}
we find the perturbative correction to the classical correlator, 
\begin{equation}
    \langle S_1 S_2 S_3 \rangle = -\frac{g}{4} \int^0_{-\infty} d \tau\, \sin B_1 \tau \cos B_2 \tau \cos B_3 \tau + \rm{perm} \ , 
\end{equation}
where $\tau = t - t'$. Evaluating the time integral by regulating the $\tau \to -\infty$ using analytic continuation, we get 
\begin{equation}\label{eq:SSSclass}
    \langle S_1 S_2 S_3 \rangle = -\frac{g}{16} \sum \bigg[ \frac{1 }{B_i \pm B_j \pm B_k} \bigg] \ .
\end{equation}
Naively, approaching the resonant pole, $B_1 = B_2 + B_3$, causes the correlator to diverge. This sharp feature is the part of the SQU proposal and is visible without actually reaching the pole directly. However, the pole itself signals the need for degenerate perturbation theory. To see this, we first note that we do not have to regulate the time integral in this way. We could instead turn on the interaction at some fixed $t = 0$. Then, we would get that $\tau$ runs from $0 \to t$, so that
\begin{equation}\label{eq:spinbi}
    \langle S_1 S_2 S_3 \rangle = -\frac{g}{16} \sum \bigg[ \frac{1 - \cos((B_i \pm B_j \pm B_k)t)}{B_i \pm B_j \pm B_k} \bigg] \ ,
\end{equation}
where the sum is over all permutations. Technically speaking, this result contains no pole, as Taylor expanding the cosine leads to a cancellation as $B_1 \to B_2 +B_2$. Therefore, we want to understand how the physical behavior close to the pole is robust to the details of the model.

In classical statistical systems, the time-dependent terms vanish as $t \to 0$ due to dephasing. Specifically, due to the nonlinear corrections, the frequencies become random and cause the oscillator terms to average to zero. This leads to the same behavior as the calculation using analytic condition, but for a physically different reason than their quantum mechanical analogues. As a simple illustration, suppose that $B_i$ was drawn for a normal distribution with variance $\sigma^2$ and mean $\bar B_i$ so that
\beq
\langle e^{i B_i t} \rangle_{B_i } =\int^{\infty}_{-\infty}  \frac{dB_i}{\sqrt{2 \pi \sigma^2}}\exp(i B_i t)\exp\left(- \frac{(B_i-\bar B_i)^2}{2\sigma^2}\right)  \propto \exp\left( -\tfrac{1}{2} t^2 \sigma ^2 \right)   \xrightarrow{t \to \infty} 0 \ .
\eeq
In this sense, we see that dephasing alone is sufficient to explain why only the time-independent contribution, Equation~(\ref{eq:SSSclass}), survives. We will see that dephasing is a general property of classical systems in Section~\ref{subsec:correlations} and Appendix~\ref{app:resonant} (See e.g.~\cite{Rioseco:2020mpc, McMillan:2008pu, Banik1, Banik2} for applications of dephasing to gallactic/stellar dynamics).

The discussion around the time-dependence gives a misleading understanding of the physical poles that appear at leading order in $g$. We have 3 spins that are precessing at rates $B_i$ that have been coupled together in a way that allows them to exchange energy. When we tune $B_1 \to B_2 +B_3$, the energy exchange corresponds to a classical resonance. Explicitly evaluating the integral when $B_1 \to B_2 +B_3$, we can see that the pole is really a signal of time evolution,
\begin{equation}\label{eq:secular}
    \langle S_1 S_2 S_3 \rangle = -\frac{g}{4} \int^t_{t_0} d \tau\, \sin (B_1+B_2) \tau \cos B_2 \tau \cos B_3 \tau 
    \supset \frac{g}{32} (B_1-B_2-B_3)(t-t_0)^2 \ . 
\end{equation}
In this regard, the divergence at $B_1 \to B_2 +B_3$ is really tied to the fact our perturbative results depends on order of limits involving $t-t_0 \to \infty$ (see also~\cite{Ballesteros:2026evr} for a similar observation). 

Fortunately, we can understand the classical evolution in the $t \to \infty$ limit using the equations of motion non-perturbatively. Defining the slow frequency as $\gamma = \omega_1 - \omega_2-\omega_3$ and slow variables $I_1 = J_1$, $I_2 = J_2-J_1$ and $I_3= J_3 -J_1$, we can eliminate the high frequencies by canonical transformations (see Appendix~\ref{app:resonant}) so that we are left with the resonant Hamiltonian, 
\beq \label{eq:spin_res_ham}
H \approx (B_1-B_2-B_3) J_1 + \frac{g}{4} J_1 (J_2-J_1)(J_3-J_1) \cos \gamma \ ,
\eeq
where $\gamma$ is the conjugate to $J_1$. The resulting equations of motion are
\begin{align}
\dot \gamma  =& (B_1-B_2-B_3) +\frac{g}{4} \left((J_2-J_1)(J_3-J_1) - J_1 (J_3-J_1) - J_1(J_2-J_1\right) \cos \gamma \nonumber\\
\dot J_1 =&-\frac{g}{4}J_1 (J_2-J_1)(J_3-J_1) \sin \gamma \ .
\end{align}
From the time evolution of $J_1$, we see that energy moves between the spins with a frequency set by $\gamma$. Moreover, notice that our solution to $\gamma$ will depend on $J_i$, but these are our random variables. As a result, we will see that statistical averages will involve averaging over this frequency, leading to similar phenomenon of dephasing. 

So, we see that the original difference between the classical and quantum vacuum fluctuations due to the resonance poles signal the need for using the resonant Hamiltonian when the resonance condition is satisfied. It is important to note that the resonant Hamiltonian keeps the classical and quantum distributions distinct, and statistical signatures survive which can be used to tell this difference. For instance, we can consider the following correlator
\begin{equation}
    {\cal O} = g( S_2^2 S_3^2 - S_1^2 S_3^2 - S_1^2 S_2^2) \ .
\end{equation}
Classically, as we show in Appendix \ref{app:CM_vs_QM_res_corr}, this correlator is deformed due to resonant exchange of energies, and tends to vanish. However, quantum mechanically, since we are in the approximately Gaussian state, this operator is non-zero. Explicitly we have 
\begin{equation}
    \langle S_2^2 S_3^2 - S_1^2 S_3^2 - S_1^2 S_2^2\rangle_{\rm CM} = 0 + {\cal O}(g)\qquad \langle S_2^2 S_3^2 - S_1^2 S_3^2 - S_1^2 S_2^2\rangle_{\rm QM} \to -\frac{1}{4} + {\cal O}(g) \ .
\end{equation}
Hence, while the resonant Hamiltonian resolves the apparent pole, it keeps the classical versus quantum signatures alive via the difference in correlators in the theory. 

\vskip 5pt
{\it Summary:} We see in this example that the quantum system in the ground state is static. The spins point in the $\hat x$ direction to minimize the energy. However, since this is not an eigenstate of $\sigma^z_i$, so we get non-trivial statistics when we measure a spin along the $z$-axis. When we add an interaction, it slightly shifts the ground state so that it becomes a superposition of the all $+\hat x$ and all $-\hat x$ states. The system remains stationary and with a large gap from the ground state to the next excited state. However, a measurement of the spins in the $\hat z$ direction will show three-point correlations suppressed by this gap.

In contrast, the classical spin is evolving in time with a non-trivial frequency. If the phases of the spins are random and independent, then statistically there will be no correlations between them at a fixed time. When we introduce a Hamiltonian that couples the spins, we are not just correlating their statistics, we are allowing energy to move between the three spins dynamically. The resulting change to the evolution depends on the differences between the frequencies, not just their sum. This gives rise to correlations that depend on these differences. Away from resonances, this will appear to generate correlations that diverge at physical frequencies. However, as we tune the frequencies to such a resonance, we have to account for the non-perturbative affect on the frequencies, which ultimately leads to a decay on the correlator associated with the on-resonance pole.

\subsection{Defining Observables}

Our central assumption is that we can describe our observations in terms of ``operators", $\O_i$, which are local quantities that evolve in time as a closed system,
\beq
\frac{d}{dt} \O_i = \sum_{j} c_{ij} \O_j  \ .
\eeq
We will additionally assume that the theory is time-translation invariant so that there is no explicit time dependence. We can assume that this evolution is generated by a Hamiltonian, but we will be careful to define what that means in both the classical and quantum context.

The label of the operators, $i,j = 1,.., M$, can run over both operators that we directly observe and additional operators needed to make this a closed system. Hidden variables theories, or theories with dissipation, will be encoded through such additional operators. We can diagonalize the evolution so that there exist a set of operators
\beq
\frac{d}{dt} \bO_i = i \omega_i \bO_i   \qquad  \bO(t) =e^{i \omega_i t} \bO(0) \ .
\eeq
Here we assumed that $\omega_j \neq 0$ so that we could trivially integrate. This assumption will be important in understanding the quantum universe signal. In principle $\omega_i$ can be complex but we are anticipating there will be constraints on $\omega_i$ arising from unitary evolution.

In terms of these operators that diagonalize the evolution, we can determine our observables by integration
\beq
\O_i(t) =  \sum_{j} \tilde c_{ij} e^{i \omega_j t} \bO_j(0)  \ .
\eeq
Despite the appearance of imaginary numbers, nothing so far distinguishes a quantum and classical theory. Now we need to break down the differences between the classical and quantum theories.

\vskip 10pt
\noindent {\bf Quantum Theory:} In the quantum description, $\bO_i$ are quantum mechanical operators evolving in the Heisenberg description
\beq
\dot \O_i =  \frac{i}{\hbar} [H , \O_i]  \to [H, \bO_j] = \omega_j \hbar \bO_j  \ .
\eeq
Here the original set of observables and the diagonal operators do not necessarily commute, $[\O_i, \bO_j ] \neq 0$. We will be specifically interested in the case where we are measuring correlations in a vacuum state. We will assume that we are in the ground state of the system, $|\Omega\rangle$ at $E = 0$ and that this state is unique.  Any operator that commutes with the Hamiltonian is therefore also in a unique state that is time independent and therefore has no variance. 

The fact that the system has a unique ground state implies that any states we can create by acting on the vacuum have higher energy. As a result, if we create a state by acting with $\bO$ on the vacuum
\beq
|\omega \rangle = \bO |\Omega\rangle \to H |\omega \rangle = H \bO |\Omega  \rangle= \hbar \omega |\omega \rangle \ .
\eeq
 Therefore, only $\omega > 0$ can create states. This is, of course, what is implemented by creation and annihilation operators in any Fock state.

 \vskip 10pt
\noindent {\bf Classical Theory:} In the classical description, $\bO_i$ are statistical variables that all commutes with all of our original observables, $[\O_i, \bO_j ] = 0$. What this means is that any observable is completely determined by a unique set of $\bO_j$. As a result, the statistical distribution of $\O_i$ for even a single $i$ means that some of the $\bO_j$ must be distributed statistically as well. Furthermore, since the time dependence is deterministic, all of the uncertainty is already present in the initial conditions, e.g.
\beq
\langle \bO_i(t) \bO_j (t) \rangle = e^{i (\omega_i +\omega_j)t } \langle \bO_i(0) \bO_j (0) \rangle \ .
\eeq
Now we note that the probability distribution can be defined in terms of a distribution $P(\omega_i; \lambda_j)$ over $\omega_i$ and a distribution over auxiliary parameters, $\vec \lambda$. Therefore, the statistics of our observable become 
\beq
\langle \O_i (t) ...\rangle =\int d\omega d^m\lambda P(\omega, \vec \lambda)  \sum_{j} \tilde c_{ij} \int dt' e^{i \omega t'}  \left( \bO(0) \ldots \right)  \ .
\eeq
Not that $\bO(0)$ depends on $\omega$ and $\vec \lambda$ so this is just one particular classical configuration at time $t=0$ and all the randomness is generated from averaging over the physical parameters.
\vskip 10pt
\noindent {\bf Constraints from Unitarity:} \hskip 5pt The final ingredient we will need is that both the classical and quantum theories are unitary. There are two important features about unitarity which constrains time evolution. One is that the probability is conserved, and the other that time evolution is invertible. These two conditions are manifest in the requirement that the Hamiltonian is self-adjoint. 

The implications of unitarity for both the classical and quantum theory can be encoded in the requirement that for every $\omega_i > 0$ there exists another $\omega_{j= -i} < 0$ (For details, see Appendix \ref{app:unitarity}). We will use the notation $\omega_{-i} = - \omega_{i}$ so that we can use $i>0$ to label the unique eigenvalues that are not fixed by unitarity.

Now we will assume that our observables are defined by a small number of operators, such as $\O_{a=1,2,3}$ and we want to calculate equal time correlators 
\beq
\langle \O_a \O_b(t) \rangle = \sigma_a^2 \delta_{ab} \qquad \langle \O_1 \O_2 \O_3(t)\rangle \ .
\eeq
Here we will assume that the two point statistics of the operators are the same in the classical and quantum theory and can be determined to be time independent. In order for this to hold, we have the following requirement for the quantum theory,
\beq
\langle \bO_{-i} \bO_j \rangle_{\rm Q}= \bar\sigma^2_i \delta_{i,j} \qquad \langle \bO_{i} \bO_{j} \rangle_{\rm Q} = 0 \ ,
\eeq
for $i>0$. This follows from our requirement that the theory has a unique ground state. In contrast, because all these operators commute in the classical theory, we must have 
\beq
\langle \bO_{-i} \bO_j \rangle_{\rm C}= \langle \bO_{i} \bO_{j} \rangle_{C} =\frac{1}{2} \bar\sigma^2_i \delta_{i,j} \ ,
\eeq
where the factor of $\frac{1}{2}$ is required to keep the two-point statistics of the observables equal in the quantum and classical theories.

Now we see that important physical difference between the quantum and classical theory: the quantum theory only has non-trivial states with $\omega_i> 0$ while the classical theory necessarily contains physical configurations with negative frequencies $\omega_i<0$. Specifically, we recall that the classical theory is just a representation of what initial conditions are present. Suppose our operator $\O_1$ contains a specific frequency $\omega_i >0$ so that $c_{i1} \neq 0$. The two point statistics therefore require that $c_{(-i)1} \neq 0$ and that $P(\omega_i|\vec \theta) > 0$ implies $P(\omega_{-i} | \vec \theta)> 0$.

\subsection{Deformations and Correlations}
\label{subsec:correlations}



In this section, we want to determine the corrections to the SQU signature on resonance. We will show that, in general, the poles need to be resummed via the resonant Hamiltonian. While this leads to time dependence and growth of correlations of initially, at late times the ensemble dephases and any time dependence in correlators averages to zero. In this way, we are able to give a complete resolution of the correlators at the poles, which was left unresolved in the original SQU proposal. 

\subsubsection*{The Resonant Hamiltonian}

Let's consider the situation where our key observables are $\O_{i=1,2,3}$ and are, to first approximation controlled by specific frequencies $\omega_{i =1,2,3}$ so that
\beq
\O_i \approx c_i e^{i \omega_i } \bO_i(0) +c^*_i e^{-i \omega_i} \bO_{-i}(0) \ .
\eeq
We will assume that we can control these frequencies sufficiently that we can arrange for $\omega_1 \approx \omega_2 + \omega_3$. Now we deform the theory by an interaction Hamiltonian, 
\beq
H = H_0 + H_{\rm int} \qquad H_{\rm int} =  g \O_1 \O_2 \O_3 \ .
\eeq
Our goal is to understand, on general grounds, how this kind of deformation alter the correlations.

First, let's consider the implications for the quantum theory. We are assuming the quantum theory is in the ground state, $|\Omega\rangle$ of the theory at all values of $g$. In perturbation theory, we have
\beq
|\Omega\rangle = |\Psi\rangle + g|\delta \Psi\rangle \qquad |\delta \Psi \rangle = \sum_{E>0} \frac{1}{E}\langle E | \O_1 \O_2 \O_3 |\Psi \rangle_{g=0} |E \rangle_{g=0} \ .
\eeq
As a consequence, we find a contribution to the correlation
\begin{labeledbox}{Quantum}
\begin{equation}\label{eq:quantum_box}
\langle \mathcal{O}_1 \mathcal{O}_2 \mathcal{O}_3 \rangle \supset g \sum_{E>0} \frac{2}{E}\left|\langle E | \mathcal{O}_1 \mathcal{O}_2 \mathcal{O}_3 |\Psi \rangle_{g=0}  \right|^2 \ .
\end{equation}
\end{labeledbox}
\noindent In situations where there is a state-operator correspondence the sum is over $E = \omega_i$ (i.e. the list of all operator frequencies), but these need not hold in general. We see that we can control the amplitude of the correlation by changing $g$ but it does not introduce any new time dependence.



Now we will consider the consequences of the same correlator in the classical theory. First, we can mirror the SQU test and solve these equation perturbatively in terms of $\bO_i = e^{i \omega_i t}\bo_i$. For example, if $H_{\rm int} \supset g \bO_{-1} \bO_2 \bO_3$, then we will find a correction to the operator, 
\beq\label{eq:operturb}
\bo_1 \supset i g\int^t  dt' e^{i(-\omega_1+\omega_2+\omega_3)t'} \bo_2 \bo_3 = -\frac{ g}{\omega_1-\omega_2-\omega_3} e^{i(-\omega_1+\omega_2+\omega_3)t}\bo_2 \bo_3 \ .
\eeq
This same term would then appear as a correction to any three point statistic
\beq\label{eq:offresonance}
\langle \O_1 \O_2 \O_3\rangle \supset -\frac{g}{\omega_1-\omega_2-\omega_3} \langle \bo_2 \bo_3 \bo_{-2} \bo_{-3}\rangle|_{g=0} + {\rm permutations} \ .
\eeq
The appearance of the pole here is identical to the one we found in in the SQU proposal, Equation~(\ref{eq:biclass}), and in the spin example, Equation~(\ref{eq:spinbi}). Like the spin example, rather than interpreting the pole as a signal of an on-shell decay, we can see that it is just a signal of a diverging time integral, as in Equation~(\ref{eq:secular}). A useful analogy here is the difference between conventional and degenerate perturbation theory in quantum mechanics. Here, we have two modes with the same frequencies, $\omega_1$ and $\omega_2 +\omega_3$, and therefore we need to resolve potential mixing non-perturbatively to correctly understand the evolution under a deformation.

In classical physics, what takes the place of degenerate perturbation theory is the time-evolution generated by the resonant Hamiltonian. The first step is the observation that the off-resonance perturbative calculation can be reproduced using a canonical transformation of the operators. Shifting $\bO_i$ and frequencies $\omega_i$ to remove the ${\cal O}(g)$ terms in the Hamiltonian introduces these correlations by a change of variables alone. This is not surprising, as the perturbative result in Equation~(\ref{eq:operturb}) is effectively a redefinition of $\bo_1$. However, this procedure, shown in Appendix~\ref{app:resonant}, only eliminates terms that already have a large time dependence, namely those that are suppressed by some linear combination of large frequencies.  However, when there exist resonant combinations of frequencies, such as $\omega_1 = \omega_2+\omega_3$, there exist operators whose time dependence vanishes as $g\to 0$ and cannot be removed in this way, e.g. 
\beq
\frac{d}{dt} \bO_{-1}\bO_2 \bO_3  = \{H, \bO_{-1}\bO_2 \bO_3  \}= 0 \ .
\eeq
In the classical theory, this shows that there are necessarily non-trivial states, $\bO_i\neq 0$, with an effective frequency $\omega_i \approx 0$. This gives a sharp contrast to the quantum theory, where the vacuum is a unique state with all excitations carrying $\omega_i > 0$.

The result is that we cannot eliminate all the terms $H_{\rm int}$ by canoncial transformations. Instead, we are left with the resonance Hamiltonian, $H_{\rm res}$, which is given by
\beq\label{eq:hprime}
H' = H + g \bO_{-1}\bO_2 \bO_3 + {\rm h.c.}\  . \equiv H + H_{\rm res} \ .
\eeq
The presence of the resonant Hamiltonian introduces non-trivial time dependence (secular terms) in the evolution of the operators $\bO_{j=1,2,3}$ due to $g \neq 0$, 
\bea\label{eq:Omixing}
{\dot \bO}_1 = i \omega_1 \bO_1 + i g \bO_2 \bO_3 \ , \quad 
{\dot \bO}_2 = i \omega_2 \bO_2 - i g \bO_1 \bO_{-3} \ ,\quad 
{\dot \bO}_3 = i \omega_3 \bO_3 -  i g \bO_1 \bO_{-2}  \ .
\eea
The factors of $i$ here arise because, in the $\bO$ basis, the Poisson bracket must take the form (for $j>0$)
\beq
\{ \bO_j,\bO_{-j} \} = i \ . 
\eeq
The key is idea is that we want a non-perturbation solution to these equations of motion for the degenerate solutions. For three modes, a complete solution can be found either numerically or in terms elliptic Jacobi functions. But it is more useful to first isolate the slow/generate modes and then solve for their evolution using secular perturbation theory via the resonant Hamiltonian. To simplify, suppose we are working on resonance so that $\omega_1 = \omega_2 +\omega_3$ and write 
\beq
\bo_i = a_i(t) e^{i \beta_i(t)} \qquad \beta_i = {\cal O}(g) \ ,
\eeq
where $a_i$ and $\beta_i$ are real. Plugging this ansatz into our equations of motion and taking the imaginary part, we find
\beq
\dot \beta_1 = g\frac{a_2 a_3}{a_1} \cos( \beta_2+\beta_3 -\beta_1) \qquad \dot \beta_{2,3} = -g \frac{a_1 a_{3,2}}{a_{2,3}} \cos( \beta_2+\beta_3 -\beta_1) \ .
\eeq
Now we notice that we can write a single equation for the relative phase $\gamma = \beta_1 -\beta_2 -\beta_3$ 
\beq
\dot \gamma = g \left( \frac{a_2 a_3}{a_1} + \frac{a_1 a_3}{a_2} + \frac{a_1 a_2}{a_3} \right)\cos(\gamma) \ .
\eeq
The key result here is that the corrected frequencies are determined by the amplitudes $a_i$. 

\subsubsection*{Resolving the Resonant Poles} 

By construction, the quantum theory in the ground state can be engineered to have time-independent correlations for generic deformations of the Hamiltonian.

For classical theory, we have seen that the existence of resonances gives rise to phases for individual fields that depends on the amplitudes
\beq
\langle \bO_1\bO_{-2} \bO_{-3} \rangle = \langle a_1 a_{2} a_{3} e^{i \gamma(t)}\rangle \ ,
\eeq
where the phase is a now a random a variable through
\beq
\dot \gamma \propto \left( \frac{a_2 a_3}{a_1} + \frac{a_1 a_3}{a_2} + \frac{a_1 a_2}{a_3} \right) \ .
\eeq
At face value, this observation is not directly useful because we have not assumed that we can measure this time dependence directly. Instead, the question is what does this time dependence imply for the correlations.

It is crucial to remember that the amplitudes $a_i$ are random numbers what we will assume are down from a smooth probability distribution $p(\vec a)$, where $\vec a$ is the list all the amplitudes. Second, it is also critical that, because $H$ is time-independent for all values of $g$, we can always decompose the operators into a diagonal basis
\beq
\O_i(t) =  \sum_{j} \tilde c_{ij} e^{i \Omega_j(g,\vec a) t} a_j(0)  \ .
\eeq
Specifically, the eigenvalues $\Omega_j$ are explicitly a function of both $g$ and $\vec a$. We can define the statistics of our model so that the frequency is independent of $\a$ at $g=0$, $\Omega_j(g=0,\vec a) = \omega_j$, but this does not hold for $g \neq 0$ due to the resonant mixing. 

Now we can use the Riemann-Lebesgue Lemma to prove that the correlations must vanish as $t \to \infty$ as follows: suppose we have an observable, $P(\a)$, that can be defined in terms of the initial conditions $a_j(0)\to \a$ and whose time dependence is decomposed in frequencies, $\Omega_j(\a) \to \Omega(\a)$ that are smooth functions of $\a$. Under the assumptions, the statistical average of $P(\a)$ vanishes at late times, 
\beq
\lim_{t \to \infty} \int d^D a P(\vec a) e^{i \Omega(\a) t}  = 0 \ .
\eeq
Here, since $\Omega_j(\a)$ is determined by $\a$, we can include the sum over $\Omega_j(\a)$ in the $\a$ integral by defining $\rho(\Omega)$ as the density of frequencies (states) as a given $\a$.  Note here that $P(\vec a) = p(\vec a)\rho(\a) F(\vec a)$ includes both the probability distribution for $\a$ and a function $F(\a)$ related to the amplitudes of the $\bO$ fields. The consequence for our model is that
\begin{labeledbox}{Classical}
\begin{equation}\label{eq:classical_box}
\lim_{gt \to \infty} \lim_{\omega_1 \to \omega_2 +\omega_3} \langle \mathcal{O}_1 \mathcal{O}_2 \mathcal{O}_3 \rangle \to \sum_{\Delta E \neq 0}\frac{c_{\Delta E}}{\Delta E} + {\rm local} \ ,
\end{equation}
\end{labeledbox}
\noindent where $\Delta E = \pm \omega_1 \pm \omega_2 \pm \omega_3$ and $c_{\Delta E}$ are constants.  We have introduced $g$ by rescaling time when $g > 0$. This is important because we, in principle, measure this behavior by changing $g$ at fixed $t$. Finally, the additional ``local" terms here are purely polynomial in $\omega_i$ (i.e.~contain only non-negative powers of $\Delta E$).

The key result here is that the non-resonant poles survive in our final expression, but the resonant Hamiltonian leads the contribution from the physical pole to vanish when we try to sit on the resonance by taking $\omega_1 = \omega_2 + \omega_3$. In this regard, we find that the appearance of additional terms in the correlation found in SQU survive off-resonance but are resolved by time evolution of the system and eventually decay due to dephasing.

The proof of the Riemann-Lebesgue Lemma is sufficiently straightforward that it is worth reviewing to see the dependence on our assumptions: given a function $f(x)$, it's fourier transform, $\tilde f(k)$ satisfies
\beq
\tilde f(k) = \int dx f(x)e^{i k x} = \int dx' f(x'+ \pi/k)e^{i k x' + i \pi} = -  \int dx' f(x'+ \pi/k)e^{i k x'} \ .
\eeq
Therefore, if $f(x)$ is a continuous function, as $k \to \infty$ we have $\tilde f(k) \to 0$.

For our example, we not only require that $P(\a)$ is smooth but also that we can change variables of integration to $\Omega(\a)$ and therefore the frequency must also be a continuous function of $\a$. This is just the inverse of our procedure for defining $\Omega(\a)$ and therefore equivalent to the assumption that $\rho(\a)$ and therefore $P(\a)$ is smooth. This procedure fails for $g = 0$ because $p(\a)$ is not smooth in the sense that $\Omega(\a) = 0$ is enforced by the time-independence of the initial statistics. However, when $g \neq 0$, the resonant mixing forces the diagonal basis of $\Omega_j(\a)$ to mix different initial frequencies. As a result, we find that all correlations will vanish as $t\to \infty$ due to the Riemann-Lebesgue Lemma.

Physically, we know that because the phase depends sensitively on the initial amplitudes $a_i$, averaging over the initial values will cause large oscillations in the correlators, which will average to zero. Mathematically, because everything is a smooth function of $\a$ at $g\neq 0$, we can change variables to integrate over $\Omega_i$ and this becomes the statement of the Riemann-Lebesgue Lemma precisely.
\vskip 10pt 
\noindent {\bf Relation to Classical Frequency Shift:} Notice that the Riemann-Lebesgue lemma applies equally to any system where the frequencies are controlled by continuous random variables. As discussed in Appendix~\ref{subapp:generalities}, this will arise under generic nonlinear deformations. Given an classical theory of a single pair of action angle variables, the addition of a angle-independent nonlinear term,
\beq
H = \omega I + g h(I) \ ,
\eeq
will give rise to an effective frequency
\beq
\dot \theta = \omega + g \frac{\partial}{\partial I} h(I) \qquad \omega_{\rm eff}(I)= \omega + g h'(I) \qquad \dot I = 0  \ .
\eeq
If we consider several such systems with random values for $I$, we will again find that correlations decay with time.

However, because the deformation by $h(I)$ does not depend on angle variables, it does not lead to time evolution of $I$. As a result, we can simply define the statistics for any given $g$ such that $I$ is drawn from a probability distribution $P(I)$ while holding $\omega_{\rm eff}$ fixed. Said differently, we can always make both $\omega$ and $I$ random with a correlation that holds $w_{\rm eff} = \omega +g h'(I)$ a constant. This is similar to what happens in our quantum theory, where we keep the system in the physical ground state as we change $g$, rather than holding the state fixed.

On resonance, we are not able to make a similar choice to hold the physical frequencies fixed. The resonant Hamiltonian changes the nature of the time evolution away from a single frequency for any values of the parameters, meaning we can see the physical implications of the resonant term for each individual realization. At a technical level, this is because $\omega I$ is independent of $\theta$ and while $H_{\rm res}(I,\theta)$ depends explicitly on $I$ and $\theta$. As a result, even if we could hold $I$ fixed, the resonant term cannot be absorbed into a constant $\omega_{\rm eff}$ because $\partial_I H_{\rm res}$ is necessarily time-dependent.

\section{ Relation to Signals of a Quantum Universe} \label{sec:relation}

The original SQU proposal worked at the level of quantum field theory in cosmological backgrounds, or in flat space with particle detectors. In this paper, we have reduced the key elements of this proposal to a purely quantum mechanical description. 

Naturally, the first question we should understand is how the effectively infinite number of degrees of freedom of QFT affects the proposal.  First, we will show how to reproduce the perturbative structure of the QFT result using coupled harmonic oscillators. Next, we will see how how introducing additional degrees of freedom leads to QFT-like decay of the correlations. Finally, we will show how the resonant Hamiltonian arises in the QFT context and the relationship to finite temperature field theory. 

\subsection{Harmonic Oscillator Construction }\label{subsec:harmonic}

 We can think of the problem as being described by 3 particles in the position basis, $x_{i=1,2,3}$ with conjugate momenta $p_i$ and a Hamiltonian
\beq\label{eq:Hp3}
H= \frac{1}{2} \sum_i \left( p_i^2 + \omega_i^2 x_i^2\right) +g x_1 x_2 x_3 \ ,
\eeq
where the masses $m_i$ are equal and we are using units where $m_i =1$. The most convenient way to solve for the time-evolution is using the interaction picture Hamiltonian evolution. We split the Hamiltonian into $H =H_0 + H_{\rm int}$ where $H_{\rm int} = g x_1 x_2 x_3$ and $H_0$ represents three harmonic oscillators. 

In the interaction picture, the position and momentum operators are given by
\bea
 x_i &=&  \frac{1}{\sqrt{2\omega_i}} \left( a_i e^{-i \omega_i t} +  a_i^\dagger e^{i \omega_i t} \right) \\
 p_i &=& \frac{i\sqrt{\omega_i}}{\sqrt{2}} \left( a_i e^{-i \omega_i t} -   a_i^\dagger e^{i \omega_i t} \right) \ .
\eea
Again, $a_i$ and $a_i^\dagger$ are either statistical variables or quantum mechanical operators. We want to show that that operators are the $\bO_i$ operators defined in Section~\ref{sec:spectra}. Specifically, in these variables (and setting $\hbar=1$) we have
\beq
H_0 = \sum_{i=1}^3 \omega_i \left( a_i^\dagger a_i + \frac{1}{2} \right) \ .
\eeq
We see that for the quantum evolution
\beq
\dot a_i^\dagger = i[H,a^\dagger]  = i \omega_i a^\dagger \qquad \dot a_i = i[H,a^\dagger]  = -  i \omega_i a \qquad [a_i,a_i^\dagger]= 1 \ .
\eeq
The same results hold for the classical theory, using the Poisson bracket $\{a_i,a_i^\dagger\} =1$. We see, therefore, that $a_i^\dagger \equiv \bO_i$ and $a_i \equiv \bO_{-i}$.

First, we can directly calculate the quantum vacuum correlators. Using conventional in-in perturbation theory, we have
\bea
\langle x_1 x_2 x_3 \rangle &=&2 g {\rm Im} \int^t dt' \langle [x_1 x_2 x_3(t'), x_{1} x_{2} x_{3}(t)] \rangle  \\
&=& \frac{g}{4 \omega_1 \omega_2 \omega_3} \frac{1}{\omega_1+\omega_2+\omega_3}  \ .
\eea
We can recover the same result from Equation~(\ref{eq:quantum_box}), labeling the energy eigenstates of the harmonic oscillator by integers $n_{i=1,2,3}$ so that $|n_1,n_2,n_3\rangle$ has energy $E= n_1\omega_1 +n_2 \omega_2 + n_3 \omega_3$ relative to the ground state. Only the state $|1,1,1\rangle$ contribute to sum so that, using the Schrödinger picture,  
\begin{align}
\langle x_1 x_2 x_3 \rangle &=\sum_{n_1+n_2+n_3>0} \frac{2g}{n_1 \omega_1 +n_2 \omega_2 + n_3\omega_3} |\langle n_1,n_2,n_3| x_1 x_2 x_3 |0\rangle |^2 \\
&= \frac{2g}{\omega_1+\omega_2+\omega_3} \left|\langle 1,1,1| x_1 x_2 x_3 |0 \rangle\right|^2 = \frac{g}{4 \omega_1 \omega_2 \omega_3} \frac{1}{\omega_1+\omega_2+\omega_3} \ .
\end{align}
We see the appearance of the total energy pole is entirely consistent with conventional perturbation theory in quantum mechanics.

When it comes to the classical theory, we notice that when $g=0$ our cubic operators contain terms like
\bea
x_1 x_2 x_3 &\supset& \frac{1}{2\sqrt{2 \omega_1 \omega_2 \omega_3} } a_1 a_2^\dagger a_3^\dagger e^{-i (-\omega_1+\omega_2 + \omega_3)t} + {\rm h.c.} 
\\
p_1 p_2 p_3 &\supset& i \frac{\sqrt{\omega_1 \omega_2 \omega_3}}{2\sqrt{2}  } a_1 a_2^\dagger a_3^\dagger e^{-i (-\omega_1+\omega_2 + \omega_3)t} + {\rm h.c.} \ .
\eea
When $\omega_1 =\omega_2 +\omega_3$, the terms on the RHS correspond to zero frequency operators. The addition of $g x_1 x_2 x_3$ to the Hamiltonian allows the operators to mix. In perturbation theory this gives rise to the physical poles
\bea
\langle x_1 x_2 x_3(t)\rangle
&\approx & i g \int^t dt' \langle [x_1x_2x_3(t'), x_{1,{\rm I}} x_{2,{\rm I}} x_{3,{\rm I}}(t)] \rangle  \\
&\supset& \frac{g}{\omega_1-\omega_2-\omega_3} \langle x_2^2\rangle \langle x_3^2 \rangle \ .
\eea
When $\omega_1 -\omega_2-\omega_3=0$, this physical pole signals a diverging time integral and and the breakdown of perturbation theory from mixing of the states (see also~\cite{Ballesteros:2026evr} for a similar observation).

We can solve for the evolution of the system as follows. Focusing only on the zero-energy contribution to the interaction picture Hamiltonian, 
\beq\label{eq:Hintp3}
H_{\rm int} \supset   \tilde g \left( a_1 a_2^\dagger a_3^\dagger +  a_1^\dagger a_2 a_3 \right) \qquad \tilde g =  \frac{g}{2\sqrt{2\omega_1 \omega_2 \omega_3}} \ ,
\eeq
we can find the evolutions equations for $a_i$ using $\{ a, a^\dagger \}_a =1 $ where $\{X,Y\}_a =\partial_a X \partial_{a^\dagger} Y -  \partial_{a^\dagger} X \partial_a Y$ and the classical equations of motion, $\dot \O =i \{ H,\O \}_a$, to find
\bea
\dot a_1 = -i  \tilde g a_2 a_3 \qquad \dot a_{2,3} =  -i \tilde g a_1 a_{3,2}^\dagger \ .
\eea
This is precisely the same system of equations we had in~(\ref{eq:Omixing}). The dyanmics of this system can also be understood in terms of the amplitudes $n_i = |a_i|^2$ by noticing that 
\beq
\dot n_1 = i \tilde g (a_1 a_2^\dagger a_3^\dagger - {\rm h.c.} )\qquad \dot n_{2,3} = -i \tilde g (a_1 a_2^\dagger a_3^\dagger - {\rm h.c.}) \ ,
\eeq
So that $\dot n_i +\dot n_{j} = 0$ for $i\neq j =1,2,3$. In this sense, we see that the conversion of $n_1 \to n_2 + n_3$ conserves $n_1 + n_2$ and $n_1 +n_3$, as expected for a decay of particles. 

In Appendix \ref{app:CM_vs_QM_res_corr}, we again use the resonant Hamiltonian in this model to construct correlators which expose the difference between the quantum vacuum and the classical theory. This is similar to the spin example considered in Section~\ref{subsec:spin}.

\vskip 10pt
\noindent {\bf Degeneracy and the Classical Limit} \hskip 5pt A key argument in this paper is that the appearance of physical poles in the classical description arises from the mixing of a large number of degenerate solutions. Normally, we think of the harmonic oscillator as having well separated energy eigenstates. So how do we see this degeneracy arising in the classical limit? As we did in Section~\ref{subsec:classical}, will introduce the particle number eigenstates, 
\beq
|n_1,n_2,n_3 \rangle = \prod_{i=1}^3 \frac{1}{\sqrt{n_i!}}( a_i^\dagger)^{n_i} |\Omega \rangle  \ .
\eeq
and recover the classical description in the limit $n\to \infty$. These states are all are energy eigenstates of the free Hamiltonian, $H_0$, with
\beq
H_0 |n_1,n_2,n_3 \rangle = \sum_{i=1}^3 n_i \omega_i |n_1,n_2,n_3 \rangle \ .
\eeq
We can see the connection to the classical energy using typical amplitude of $a_i^\dagger$,
\beq
|\bar a_i|^2 = \langle n_i | a_i a_i^\dagger |n_i \rangle = n_i  \quad \to  \quad E_i = |a_i^2| \omega_i  \ .
\eeq
In this sense, we should be able to use either quantum or classical intuition to described the degeneracy.

To see the degeneracy in the quantum description, let's again suppose we have a situation where $\omega_1 = \omega_2 + \omega_3$. With this assumption, we can construct exactly degenerate quantum states via
\begin{align}
H_0 |n_1+q,n_2-q,n_3-q \rangle  =&  (n_1\omega_1 + n_2 \omega_2 +n_3 \omega_3 + q (\omega_1 -\omega_2-\omega_3) ) |n_1+q,n_2-q,n_3-q \rangle \nonumber \\ 
=&(n_1\omega_1 + n_2 \omega_2 +n_3 \omega_3)  |n_1+q,n_2-q,n_3-q \rangle \ .
\end{align}
We see that the energy is identical for all values of $q\leq n_i$. The consequence is that in the classical limit, where we take $n_i \gg 1$, there are a large number of degenerate states that can mix when the theory is deformed. 

\subsection{Hidden Variables and Many Degrees of Freedom}

In the conventional discussion of Bell-tests, one focuses on a classical theory of hidden variables as the alternative to quantum mechanics. The concept in that context is that the perceived randomness is due to additional classical degrees of freedom that are not measurable directly.

The possibility of a hidden variables theory is already implicit in our general construction. To recover a hidden variables theory, we assume the sector of ``visible" operators $\O_{i=1,..m}$ are quantities with measurable statistics in both the classical and quantum theory. The hidden variables are the set of additional operators ${\cal O}_{i >m}$ are not resolved and may have different statistics in the two theories. For example, we might only measure the position of a single degree of freedom $\O_1 = x$ in a classical or quantum theory with a Hamiltonian, 
\begin{equation}
    H = H(x, p, u_i, v_i) \ ,
\end{equation}
that couples the particle to $N$ conjugate pairs of the hidden variables, $u_i, v_i$. Here $p$ is the conjugate momentum to $x$ but we do not measure its statistics directly. The evolution of our visible quantities is now controlled by these hidden variables via the equations of motion \ , 
\begin{equation}
    \dot{x} = f_1(x(t), p(t), u_i(t), v_i(t)) \qquad \dot{p} = f_2(x(t), p(t), u_i(t), v_i(t)) \ .
\end{equation}
Assuming that we have a confining potential, the motion of $x(t)$ will be quasiperiodic and will have $N+1$ fundamental frequencies. Thus, we can again break up $x(t)$ into a sequence of $N+1$ Fourier modes,
\begin{equation}
    x(t) = \sum_{i= 1}^{N+1} (a_i e^{i \omega_i t} + b_i e^{-i \omega_i t}) \ ,
\end{equation}
where the $\omega_i$ depend on the interactions between the particle and the hidden variables. This is just a realization of the above construction where $\O_{i>1} = \{p,u_i,v_i\}$ are not measured and $a_i$ and $b_i$ are the frequency eigenstates, $\bO$, whose initial values depend on the details of the hidden variables.

For small or moderate $N$, we described the classical evolution by diagonalizing the list of operators. The consequence is our observable operators are not eigenstates of the evolution and become a linear combination of nearly degenerate states. 

In the limit of a large number of hidden variables, $N\to \infty$, the behavior of our observables is better understood without diagonalization. The reason is that with $N$ degenerate operators, our original operator is a vector in a very high dimensional space are quickly becomes mixed with a large number of unobservable variables. Therefore, it is useful instead to study the evolution of just a single observable, $\O_1$, coupled to a list of operators, $U_k$, with energies $\omega_k$ with 
\beq
H_{\rm int}\supset \sum_k \left( c_k \Pi_{1} U_k + c_k^* \O_1 \Pi_{k} \right) \ ,
\eeq  
where $\{{\cal O}_1, \Pi_1\} = \{U_k, \Pi_k\} = 1$. Keeping only the linear operators, one can write this as a familiar example of operator mixing in time-dependent perturbation theory
\beq
{\dot \O}_1(t) = - i 
\omega_1 \O_1 + \sum_k i c_k U_k \qquad {\dot U}_k(t) = - i \omega_k U_k + i c^*_k \O_1 \ .
\eeq
Solving for $U_k$ in terms of $\O_1$ we get
\beq
\O_1(t) = - \int dt' \sum_k |c_k|^2 e^{i \omega_k(t-t')} \O_1(t') \ .
\eeq
This is a general (formal) solution, but in the limit of large $N$ we can use the Wigner-Weisskopt approximation (see e.g.~\cite{LeBellac_2006}) to see that it leads to exponential decay. The key idea is that in the limit of $N\to \infty$ we should replace the sum over $k$ with an integral over a density of states, $\rho(\omega)$, such that 
\beq
\O_1(t) = - \int^t dt' \int d\omega \rho(\omega) |c_k|^2 e^{i \omega(t-t')} \O_1(t') \ .
\eeq
Now we take an exponential ansatz $\O_1(t) = \O_1(0)e^{i \Omega t}$ where $\Omega = \omega_1 + \delta + i \Gamma$ is complex. We then have
\beq
i\Omega = - \int d\omega \rho(\omega) |c(\omega)|^2 \frac{1}{i (\omega - \Omega)} \ . 
\eeq
Solving perturbatively in $\Gamma$, and using
\beq
\lim_{\epsilon\to 0} \frac{1}{x + i \epsilon} = {\mathbb P} \frac{1}{x} - i\pi \delta(x) \ ,
\eeq
we find 
\beq\label{eq:Gamma}
\Gamma  = \pi \rho(\omega_1) |c(\omega_1)|^2 \qquad \delta ={\mathbb P}  \int d\omega \rho(\omega) |c(\omega)|^2 \frac{1}{i (\omega - \omega_1)} \ . 
\eeq
In this precise sense, we see that the limit $N\to \infty$ gives us the usual exponential decay result.

\subsection{Origins in Quantum Field Theory}

The original SQU test was based on the structure of correlators of a field $\phi(\x,t)$ defined in Equation~(\ref{eq:phiQSU}). Having now understood the structure of the correlations at the level of quantum mechanics, we can now return to quantum field theory to identify the same structures. 

An important difference between the classical and quantum theory can be traced to the resonant Hamiltonian. Given a set of waves with dispersion $\omega  = c_s k$, one can isolate the resonant Hamiltonian that corresponds only to the terms that conserve energy~\cite{PhysRev.127.1918,Zakharov1968,Krasitskii_1994,LVOV199738,Arkani-Hamed:2026rsz}. For example, given our field $\phi(\x,t)$ written in terms of $a_\k$ and $a_\k^\dagger$, we can write
\beq\label{eq:resonantH}
H_{\rm R} = \int \frac{d^dk_1 ..d^dk_n d^d p_1 ..d^d p_m}{(2\pi)^{d(n+m)}} T(\k_i, \p_j) a^\dagger_{\k_1} .. a^\dagger_{\k_n} a_{\p_1} .. a_{\p_m} (2\pi) \delta(\sum k_i - \sum p_j)  \ ,
\eeq
where $m,n \geq 1$. The resonant Hamiltonian defines the classical scattering of waves and is related to the on-shell action. At tree level, we can use this to compute the leading $n+m$ point function
\beq
\langle \phi(\k_1,t) .. \phi(\p_m,t) \rangle  = \frac{T(\k_i, \p_j)}{2^{(n+m)/2}\sqrt{ k_1..k_n p_1..p_m}} \Delta t \ .
\eeq
We see that the coefficient $T(\k_i,\p_j)$ is proportional to the scattering amplitude in precisely the way we would have anticipated for the physical pole, $\Delta t \leftrightarrow 1/(\sum_i k_i -\sum_j p_j)$. The utility of the resonant Hamiltonian in this context is very much the same as our proposal; it calculates the leading secular behavior and simplifies the calculation of the non-perturbative time evolution. Specifically, we notice that the resonant Hamiltonian in Equation~(\ref{eq:resonantH}) is just a generalization of the Hamiltonian we defined in Equation~(\ref{eq:Hintp3}). Of course, this is the same as our general deformation, Equation~(\ref{eq:hprime}), if we take $n+m=3$ and identify $\bO_{-1} = a^\dagger_{\k_2}$ and $\bO_{2,3} = a_{\p_{2,3}}$.

In the language of QFT, our general result is the statement that the classical theory can be expressed in terms of a resonant Hamiltonian. The resonant Hamiltonian gives rise to the secular growth and eventual decay of the correlations. However, the resonant Hamiltonian annihilates the vacuum and therefore does not affect quantum vacuum correlations. 

These properties are well-known in the context of thermal field theory~\cite{Laine:2016hma,Delacretaz:2026owo}. At finite temperature, single particle states are unstable due to loop corrections that produce thermal broadening (a finite width). For example, in $\lambda \phi^4$, a single particle $\phi$ state is stable, but when we introduce $T>0$ the propagator acquires a width~\cite{Aarts:1996qi}
\beq
G^R(\omega,\vec k) = \frac{1}{-(\omega+i \epsilon)^2 +p^2 +m^2} \to \frac{1}{-(\omega+i \Gamma)^2 +p^2 +m^2}
\qquad \Gamma  =\frac{\lambda^2 T^2}{1536 \pi m} \ .
\eeq
This result remains finite as $m\to 0$ due to the thermal mass correction that leaves $m_{\rm eff} \to\lambda T^2/24$. Similar finite temperature widths arise for fermions at the Fermi surface~\cite{Chubukov_2003}.

We emphasize here that there field theory results are precisely what we would expect for a large number of degrees of freedom, as we found in Equation~(\ref{eq:Gamma}). When we observe only a small fraction of the available degrees of freedom, the information rapidly decays. The important distinction with our general results is that they hold even for a few degrees of freedom where we observe all of the available modes. The decay of correlations on resonance is not the result of information moving from visible to invisible degrees of freedom, but due to dephasing of the frequencies that depend on random variables.

\section{Comparison with Other Quantum Signals}\label{sec:compare}

There are many approaches to testing the quantum nature of reality. Naturally, one would like to understand the relationship between these different proposals. In the context of this paper, it is nature to compare to tests that do not involve a measurement of two non-commuting variables, but instead are rooted in the difference between classical and quantum evolution. We will focus on three specific examples: sub-Poissonian noise, Wigner negativity, and quantum speedups of quantum search algorithms.

However, this subject is clearly much broader. The subject of quantum speedups, or quantum advantages, enabled by quantum computers is essentially the same kind of test in our proposal: we could tell if a computer was classical or quantum by its ability to answer certain questions in a given time with fixed resources. Like the SQU proposal, quantum speedups arise because some property of time evolution of the state of the computer cannot be achieved with any classical evolution. The examples in this section are meant to compare and contrast specific results with the hope of identifying how such a broad connection could work in detail.

\subsection{Sub-Poissonian Noise and Quantum Optics}

A natural point of comparison for our SQU test is graviton detection. As in cosmology, one similarly has a field whose existence is unambiguous but whose quantum nature is not directly measurable. In this case, it has also been proposed that the sub-Poissonian statistics of a graviton detector could be used to establish its quantum origin~\cite{Mandel:77,PhysRevLett.49.136,Kanno:2018cuk}. Our goal here will be to show that, despite these similarities, the tests are not equivalent\footnote{We thank Dan Carney for raising this specific question, which motivated some of this work.}.

We will consider a simple model of a graviton detector with photons excited by fluctuations in the metric to infer the presence of the gravitational field. The Hamiltonian of the coupling between the photons, $a,a^\dagger$ and the gravitational field, $b,b^\dagger$ will taken to be
\beq
H=ig(a^\dagger b-ab^\dagger) \ ,
\eeq
where $g$ is a parameter dependent on the detector. Here $a$ and $b$ are modes in the same sense as Equation~(\ref{eq:phiQSU}) for some specific fourier mode $\k$. We will assume the detector itself, $a,a^\dagger$ are quantum an obey the quantum statistics, Equation~(\ref{eq:qstats}), while $b,b^\dagger$ could be either classical or quantum.

Following the setup described in~\cite{Carney:2023nzz,Carney:2024dsj}, we imagine ultra-fast monitoring of the photon system. We start in the ground state, $a |0\rangle =0$. When a single excitation of a photon occurs, we add to our photon count $N$ and reset the photon system to the ground state. 

Without monitoring the system, the probability of exciting $n$-photons from the ground state would be 
\beq
p(n)=\braket{\beta|K^\dagger_n K_n|\beta}\ ,
\eeq
where 
\beq
K_n=\braket{n|e^{-i\int dtH(t)}|0} \ ,
\eeq
is the Krauss operator. For the moment, we will assume the gravitational field is a coherent state $|\beta \rangle$, defined by $b|\beta \rangle = \beta |\beta\rangle$. Now if we monitor the system with a cadence $\Delta t$, then to order $\Delta t^2$ the evolution in controlled by a single photon excitation, 
\beq
K^\dagger_nK_n\approx(1-g^2\Delta t^2b^\dagger b)\delta_{n,0}+g^2\Delta t^2b^\dagger b\delta_{n,1} \ .
\eeq 
Because we reset the state to the ground state after a time $\Delta t \to 0$, we will maintain the system in $n=0$ or $n=1$ over a long period of time, effectively using the quantum Zeno effect~\cite{Misra:1976by,Facchi:2008nrb}.

For each time step, there is a probability 
\beq
p(1)=g^2\Delta t^2\braket{\beta|b^\dagger b|\beta}=g^2\Delta t^2|\beta|^2
\eeq
that we find one click in the detector for mode, otherwise we find overwhelmingly nothing. \\
By itself, this result tells us nothing about the nature of $b$, which could be a purely classical probability amplitude. However, when this occurs, we register the measurement by updating the total photon count, $N \to N+1$, and continue monitoring. It is instead the statistics of the total count $N$ that we will be interested in. 

Now monitor the detector for a time interval $T= M \Delta t$, imagining that we could repeat this experiment many times. Assuming that $p(1)$ is a constant, then average number of photons detected after a time $T$ will be give by
\beq
\braket{N}=M p(1)=\frac{T}{\Delta t}p(1)=\eta\phi T \ , \qquad \phi_\beta \equiv \braket{\beta|b^\dagger b|\beta} = |\beta|^2 \ , \qquad \hat \phi = b^\dagger b  \ ,
\eeq
where $\eta$ is the detector efficiency and \(\phi_\beta\) is the flux of the incoming gravitational wave. We have also defined the number operator, $\hat \phi$ as the comparison of $\hat \phi$ and $\phi_\beta$ will be crucial in our understanding of the photon statistics.  The assumption that the rate is a constant (or $\phi_\beta$ is constant) is precisely what gives rise to Poisson statistics, and therefore we can determine the full probability distribution
\beq
p(N)=\frac{[\eta\phi T]^N}{N!}e^{-\eta\phi T} \ .
\eeq
Under this strict assumption, all statistics of $N$ will match the Poisson statistics.

However, we can easily generalize beyond Poisson under circumstances where $\phi_\beta$ can be treated as a random variable. In the classical description, $b = \beta$ and $b^\dagger=\beta^*$ are draw from a distribution $P_{\rm cl}(\beta)$ and we compute statistics of $\phi_\beta = |\beta|^2$  In this model, the statistic of $N$ require marginalization over $\phi_\beta$ so that 
\beq
p_c(N)=\int db P_{cl}(\beta)p_1(N,\beta) \ .
\eeq
In this case, the variance of N is given by 
\beq
\langle (N-\langle N \rangle )^2\rangle =\braket{N}+\eta^2T^2[\braket{\phi_\beta^2}-\braket{\phi_\beta}^2] \geq \braket{N} \ .
\eeq
We see in the classical probabilistic description where $b$ is drawn from a distribution, $P_{cl}(b)$, the variance is always larger than purely Poissonian expectation.

If $b$ and $b^\dagger$ are operators in a quantum system described by a density matrix $\rho$, we can use the coherence state description to arrive at a similar formula. Using the Glauber–Sudarshan P representation~\cite{Sudarshan:1963ts,Glauber:1963tx}, we can write the density matrix in terms of a basis of coherent states as, 
\beq
\rho = \int d^2\beta P_Q(\beta) |\beta \rangle \langle \beta | \qquad b |\beta \rangle = \beta |\beta \rangle\ ,
\eeq
where $P_Q(\beta)$ is a pseudo-probability. Here we used the defining feature of a coherent state, namely that it is an eigenvector of the annihilation operator, $b$, with complex eigenvalue \(\beta\). 
Given that $\phi$ is just a number in any coherent state $\beta$, we can again determine the variance in $N$ just by integrating over $\beta$ to find
\beq
\langle (N-\langle N \rangle )^2\rangle =\braket{N}+\eta^2T^2\int d^2\beta P_Q(\beta)[\phi_\beta-\braket{\phi_\beta}]^2 \ . 
\eeq
Unlike the classical description, $P_Q(\beta)$, is not strictly positive and therefore it is possible to have a sub-Poissonian variance in the quantum description. To understand the origin of the possible negativity, we must notice that the replacement
\beq
\langle \beta | f(\phi_\beta) |\beta \rangle \to f(|\beta|^2)
\eeq
that was used to evaluate $\phi$ requires that operators are all normal ordered, $f(\phi)\equiv : f(\hat \phi):$. This is because $|\beta\rangle$ is an eigenvector of $b$ but not $b^\dagger$. The result is that the apparent ``variance" of $\phi$ is 
\begin{align}
\int d^2\beta P_Q(\beta)[\phi_\beta-\braket{\phi_\beta}]^2 &=\braket{:\hat \phi^2:}-\braket{\hat \phi}^2 \\ 
&=\langle (\hat\phi-\braket{\hat \phi})^2\rangle-\braket{\hat \phi}\le \langle (\hat\phi-\braket{\hat \phi})^2\rangle \ ,
\end{align}
where we recall that $\phi_\beta = \langle \beta  | \hat \phi |\beta \rangle  =|\beta|^2 $ and $\hat \phi = b^\dagger b$. In this sense, we see that the variance of pseudo-classical variable that controls the detector response is smaller than the true variance of the operator $\hat \phi$.

It is tempting to make a connection here between the QSU test given that this result appears to depend sensitively on the commutator
\beq
\hat \phi^2 = b^\dagger b b^\dagger b = b^\dagger b^\dagger b b + [b,b^\dagger] b^\dagger b = :\hat \phi^2: + [b,b^\dagger] \hat \phi \ .
\eeq
At face value this would seem to offer a test similar to the QSU, as it doesn't directly require a measurement of non-commuting observables (as we are just counting values of $N$) but the statistics are influenced by a non-zero commutator associated with the time evolution generating a non-Gaussian distribution in terms of $\phi$.

To better understand the connection, it is useful to work instead in the number basis of $b$. After all, the SQU test as directly about particle creating and annihilation which does not have a clear definition in the coherent state basis. Suppose that we start in a number eigenstate of gravitons $|M\rangle$. Under time evolution, in a time $\Delta t$, the state evolves to 
\beq
|0\rangle_a \otimes |M\rangle_b \to  (1- g^2 \Delta t^2 M)|0\rangle_a \otimes |M\rangle_b  + i g \Delta t M |1 \rangle \otimes  |M-1 \rangle \ ,
\eeq
where we used $b^\dagger |M\rangle_b = M |M-1\rangle$. Now if we detect a photon, we reset the state of the photons to the ground state, but we can't alter the state of the graviton, so the resetting operator is
\beq
|1 \rangle \otimes |M-1 \rangle \to|0 \rangle \otimes |M-1 \rangle \ .
\eeq
So we see that the probability of detecting the first photon is $g^2 \Delta t^2 M^2$, but the second will be $g^2 \Delta t^2 (M-1)^2$, and so on for more photons.

Notice that because we only detect one photon at a time, we are always acting on the state $
|M\rangle$ with $b$, lowering the state. This is why the statistics of $N$ are controlled by $: \hat \phi^n:$ rather than $\hat \phi^n$. Since only the normal ordered operators ever appear in the statistics of $N$, we never observe the statistics of $\hat \phi$ directly.

Since the probability is not the same for any time interval $\Delta t$, this does not obey Poisson statistics. Instead, the probability of detecting $n$ photons at time $t+\Delta t$ for a state with $m$ gravitons obeys the difference equation
\beq
P(n,m,t+\Delta t) = (1- g^2 \Delta t^2 m) P(n,m,t) + g^2 \Delta t^2 (m+1)P(n-1,m+1,t) \ .
\eeq
That this leads to sub-Poissonian noise is easy to see in the long time limit. If we start with $M$ gravitons, we will eventually reach a state where we have detected exactly $M$ photons with zero variance. 

In the end, we found a purely classical evolution equation for the probability distribution which reproduces the exact statistics. The appearance of the commutator in the coherent state description is only needed to explain the discrete statistics within a superposition of coherent states. However, this is not a fundamental feature of what we observe statistically, only a manipulation that is needed to explain that observation in a particular (coherent-state) basis. The same exact calculation in a different (particle number) basis shows no need for the commutator. 

Finally, it is important to distinguish the quantization of the gravitational field with a Bell test. The idea that gravitational waves only come as discrete particles is a profound consequence of sub-Poissonian noise. Indeed, we need to use states with a discrete number of gravitons to recover the correct statistics. However, discreteness can arise purely classically and does just does not additionally require that the particles are created by non-commuting operators.

What is similar about sub-Poissonian noise and the SQU test is that both exploit the fact that the quantum vacuum can distinguish $a$ and $a^\dagger$. This feature, along with ultra-fast monitoring, prevents excited photons from producing gravitons, forcing the graviton numbers to decrease when it is discrete. This is the ultimate origin of the sub-Poissonian signature. However, it is the fact that we know the photons are in the quantum vacuum that makes this test possible. The gravitons, on the other hand, must be in an excited state where both $b$ and $b^\dagger$ have a non-vanishing effect.





\subsection{Negativity of the Wigner Distribution}

The Wigner distribution offers another example of a pseudo-probability distribution where negative values are tied to quantum phenomena.
Concretely, in a quantum system in the position basis $\x$, with density matrix $\rho(\x,\x')$, we can define the Wigner distribution as
\beq
W(\x,\p)
=
\frac{1}{(2\pi)^n}
\int_{\mathbb{R}^n} d^n y
e^{i \p \cdot \y}\rho(\x-\tfrac{1}{2}\y,\x+\tfrac{1}{2}\y) \ .
\eeq
The Wigner distribution behaves like a probability distribution on phase space, $(\x,\p)$. However, since only the diagonal elements of $\rho$ are required to be positive for the quantum theory to be well defined, the Wigner distribution does not respect positivity. 

The Wigner distribution is a natural object of interest for tests of quantum mechanics in cosmology~\cite{Ireland:2026txt}. The fluctuations outside the horizon act as a quantum random walk, the quantum generalization of stochastic inflation~\cite{Vilenkin:1983xq,Starobinsky:1986fx}, can naturally be described by a Fokker-Planck equation for the Wigner distribution~\cite{LaFlamme:1990kd,Nambu:1991vs,Li:2025azq,Green:2025hmo}.

As a test of quantum origin of structure, the Wigner distribution is particularly appealing because of Hudson's Theorem~\cite{1974RpMP....6..249H}. It proves that for a pure state, $\rho= |\psi \rangle \langle \psi|$, a positive Wigner distribution,
\beq
W_\psi(x,p) \ge 0 \quad \forall (x,p) \in \mathbb{R}^{2n} 
\eeq
is possible if and only if $\psi(x)$ is a Gaussian function up to translations. This suggests a powerful connection between non-Gaussian statistic and quantum mechanics. In particular, one might confuse the QSU test with Hudson's theorem as both suggest non-Gaussian statistics are what distinguish quantum from classical~\cite{Howl:2020isj}.

Unfortunately, as an observational tool, Hudson's theorem does not offer a novel test of the quantum origin of structure directly. Specifically, if we only make measurements in the $x$ basis, we only measure properties of the marginalized distribution,
\beq
P(x) =\int dp W(x,p) \geq 0 \ ,
\eeq
which is still indistinguishable from a classical probability. Moreover, if $x$ is non-Gaussian, it only implies that $W(x,p)<0$ if $x$ is in a pure state. It is, of course, still possible to generate the same statistics classically from a mixed state (after all, a classical pure state would have trivial statistics).

A second issue is that, even in a pure state, the negative regime is necessarily in the highly non-Gaussian regime. As a result, one cannot determine the form of the Wigner negativity perturbatively. It was shown in~\cite{Ireland:2026txt} that for some models the Wigner function can still be calculated non-perturbatively to determine the structure required by Hudson's theorem. It remains unclear how this can inform any quantum signals that can be measured in the $x$ basis alone, and whether this perspective is distinct from the SQU test.


\subsection{Quantum Walks and Quantum Search}

Quantum speedups, or quantum advantages, present a useful model for how time-evolution might serve as a quantum test. The idea is that since specific calculations, like prime factorization~\cite{Shor:1994jg} or search~\cite{Grover:1997fa}, are very likely to be faster on a quantum computer than any classical computer, we can infer that a computer that produces these results must evolve the system in a way that has no classical analogue. As we will stress, however, it is important to define the comparison between classical and quantum computation in a way sufficiently general to include all possible time evolution. 

There is currently no general theory of speed-ups. While algorithms such as Shor's are generally believed to be robust examples of exponential speedups, there is no classification of speedups in which to place the SQU test. However, there are several classes of examples that are weaker quantum tests than SQU, and we will highlight in this section how and why those examples do not meet the same standards as a test of quantum versus classical time evolution. 

\subsubsection*{Correlation Based ``Speed-ups"}

One of the most common sources of quantum speedups arise from quantum induced correlations. The correlations explain the behavior and, as we will see, can be achieved classically. The origin of the apparent speed-up is instead related to the control over noise is believe to be possible in quantum systems and is often assumed to beyond the control of a classical system with noise.

The (quantum) random walk~\cite{Kempe:2003vul} provides a simple illustration of the challenge of using ``speed-up" as a Bell-like test. This example is a useful starting point given that the classical models of random walks are extremely well studied, while quantum walks have played a significant role in quantum search algorithms. Classically, an unbiased random walk diffuses at a rate $x \propto \sqrt{t}$.
The simplest quantum analogue of a random walk is called the Hadamard walk. Here, the walker is a quantum system coupled to a qubit, which acts as a coin. At each interval, we apply the following transformation
\begin{equation}
    \ket{\psi}_w \otimes \ket{\psi}_c \to e^{i p \sigma_z} \otimes H \ket{\psi}_w \otimes \ket{\psi}_c \ .
\end{equation}
For instance, $\ket{x} \otimes \ket{\uparrow} \to \frac{1}{\sqrt{2}}\ket{x+1} \otimes \ket{\uparrow} + \frac{1}{\sqrt{2}}\ket{x-1} \otimes \ket{\downarrow}$. It was shown in \cite{Kempe:2003vul} that such a walker diffuses ballistically, i.e., we have $\langle x \rangle \propto t$. 

It is tempting to describe this difference as a quantum speed-up. After all, the walker reasons a given value of $x$ faster with a well-chosen Hadamard walk than it would with an unbiased classical walk. However, there are many classical systems in nature which diffuses ballistically as well. So a measurement of ballistic diffusion cannot be taken as a quantum signal. For example, a classical random walk where all $N\propto t$ steps are correlated can easily move a distance $x\propto t$ by moving preferentially in one direction.

The issue with jumping to these kinds of conclusions is that we compare classical and quantum walks while giving them different rules for evolution. Note that classically, we defined a random walk via a coin flipping experiment, where at each step we flip a new coin. However, quantum mechanically, we defined the Hadamard walk to be the walk generated by using just one qubit, which is correlated at each step. Thus, in fact the Hadamard walk even fails to be Markovian. The entire reason the quantum walk gets the speedup is because of correlations built in via reusing the same coin. This immediately tells us that if we rather defined a classical walk which had correlations allowed as well, we could achieve the quantum speedup as well. On the flip side, a quantum random walk where a new coin is used every time step is well known to reproduce the classical behavior \cite{Brun:2003wby}. 

The comparison of random walks is also equivalent to the comparison of the classical and quantum Cram\'er-Rao bound. While there is a similar comparison of $\sqrt{N}$ versus $N$ scaling, with $N$ is the number of systems being measured. However, the quantum and classical Fisher matrices are identical for measurements in a single basis. Moreover, for multi-parameter measurement, the quantum case is less precise due to the uncertainty principle.

\subsubsection*{Defining a Common Basis}

The second challenge of comparing quantum and classical algorithms is that in a quantum system, we must define the time evolution so that it can work in any basis. By contrast, a classical system has a specific basis. For example, position and momentum are physically distinct and some quantum advantages maybe simply reflect the ability for the quantum system to solve the problem in another basis where it is easier. However, the classical system, if allowed to access that basis, may have a similar performance. Thus, this is a weaker test than an SQU test in the sense that we do not assume the quantum and classical description are required to be defined in particular bases.

A simple illustration is the Deutsch–Jozsa algorithm~\cite{Deutsch:1992idk}.  Suppose we have a function $f(\{0,1\}^n) \to \{0,1\}$ that is either always the same $f(\vec x) = 1$, or $f(\vec x) = 0$, for all $\vec x \in \{0,1\}^n$ or balanced so that $f(\x)$ is equally likely to be 0 or 1. For the classical algorithm, if you want to know the answer with absolute certainty, then you need to make at least $n/2 +1$ measurements (you need to try enough points to find one that doesn't match in the balanced case). For the quantum algorithm you only need 1 measurement. We do this as follows: we define a super-position for  $\x \in \{0,1\}^n$  

The basic idea is that you make a state that is a super position over $\x$:
\beq
|\psi \rangle = \sum_\x |\x \rangle \left(|0\rangle - |1\rangle \right) \ .
\eeq
Then you act with the unitary transform 
\beq 
|\psi \rangle \to \sum_\x |\x \rangle \left(|0 + f(\x)) \rangle - |1+ f(\x)  \rangle\right) = \sum_\x (-1)^{f(\x)} |\x \rangle \left(|0\rangle - |1\rangle \right) \ .
\eeq
At this point, it essentially clear what will happen: if $f(\x) = {\rm constant}$ then this state is just
\beq
|\psi \rangle  = c  \left(|0\rangle - |1\rangle \right) \otimes \sum_\x  |\x \rangle \ ,
\eeq
where $c$ is a constant. If it is not a constant, it will be in some other state.

The central claim here is that we cannot do this with a classical system with only 1 call to $f(\x)$, but what we really mean is that $f(\x)$ is a linear function on a Hilbert space in the quantum theory but is a scalar function on $[0,1]^n$ for classical system. If we instead defined $f(\x)$ as a matrix $F(\X)$ and defined $\X$ to be 2n dimensional vector, where each $\x \in [0,1]^n$ is assigned a coordinate direction in the 2n dimensional space $X \leftrightarrow (\x_0, \x_1,.., \x_{2n})$, ie so that $\X = (1,0,...,0)$ is the same as $\x_0$ (some specific element of $[0,1]^n$). The matrix $F$ is diagonal so that $F_{ii} = f(\x_i)$. We could determine if $f(\x)$ is a constant with a single call to $F$ using the vector $\X_1 = (1,1,..,1)$ and calculating 
\beq
a=\frac{1}{2n} \X_1^{T} F \X_1 = \frac{1}{2n} \sum_{\x} f(\x) \ .
\eeq
If $a =0$ or $a=1$ then, $f(\x)$ is constant, otherwise $f(\x)$ is not a constant.

This classical description does require exponentially more resources than its quantum counter-part, we note that for a Bell-type test there was no restriction on the dimensionality of the space of states. Moreover, we can see that the number of states is only needed to mimic the ability to call the function $f(\x)$ as a linear operation in any super-position. Even this is not required for a classical algorithm to reproduce the behavior of the Deutsch–Jozsa algorithm classically. Instead, we only need to be able to call $f(\x)$ as a single superposition.

To make contact with a wider range of quantum algorithms, let us look at this problem through the lens of the fourier/momentum representation. Suppose that instead of calculating $f(\x)$, we are able to calculate its discrete fourier transform
\beq
\tilde f(\p) = \sum_{\x} e^{i \pi \vec p \cdot \x} f(\x) \ ,
\eeq
where $\p \in \{0,1\}^n$. In order to determine if the original function $f(\x)$ is balanced, we only need to know the value of $\tilde f(\p)$ at $\p = 0$,
\beq
\tilde f(0) =\begin{cases}
\pm n , & \text{if } f(x) = \pm 1 \qquad \forall \x, \\
0, & \text{if } f(\x) \qquad \text{is balanced}.
\end{cases}
\eeq
We see that we can define a classical scalar function that we can call once to determine if $f(\x)$ is balanced or constant. The unique power of quantum computers for this kind of problem is that the operator $\hat f$ can be evaluated in any basis, whereas a classical operator with the same flexibility requires using a large vector space.

This behavior is exhibited by a large number of algorithms including the exponential speedup provided by the quantum search algorithm on a random graph~\cite{Childs:2004tgh,Childs:2008vkn,Lovett:2010uil}. Treated as a classical walk, a random walker must proceed through the graph one site at a time. If the graph has distinct features at a single site, it can take exponentially long for the random walk to find its target/exit. However, the graph has some useful global structure, a quantum algorithm can work on a linear combination of sites to determine exits quickly. Although these algorithms may or may not be exploiting some uniquely quantum aspects of the system, it is also clear that these constructions most easily demonstrate a speedup by limiting the kinds of classical problems that can be used in comparison. It is therefore not a given that quantum search speedups generically indicate uniquely quantum behavior. For example, it was shown in~\cite{Brown:2019jvs} that the Grover search algorithm~\cite{Grover:1996rk} is mathematically equivalent to a classical mechanics problem.

\subsubsection*{Summary and Future Directions}

The connection between quantum advantage and tests of quantum time-evolution suggests a potentially exciting synergy between research in quantum computing and cosmological Bell tests. However, as we have seen, the simplest individual examples of quantum speedups do not align with the statistical Bell-type tests we discussed here. Yet, this is more of a reflection of the incomplete understanding we have of both topics. A general theory of quantum advantage has yet to be developed, and many of the proposed advantages have not been rigorously proven. Attempts to better understand the origin of quantum advantages, such as~\cite{Thio:2025hbc}, mirror ideas that have developed in the context of cosmic Bell test~\cite{Ireland:2026txt}. This suggests that there are compelling opportunities at the intersections of these fields that will emerge as we better understand how quantum evolution (in nature or in a computer) differs from its classical counterpart.

\section{Conclusions}\label{sec:conclusions}

The quest to understand the quantum nature of reality remains one of the most critical questions in physics. While Bell's inequalities offer a definitive test in simple systems, the question of how quantum mechanics manifests itself in complex systems remains harder to pin down. The goal of proving quantum speed-ups, or quantum advantages, for quantum algorithms with real world applications give concrete and critical examples of where our current understanding is insufficient.

The broad hope in many examples, from cosmology to quantum computing, is that time evolution itself could play the role of a Bell test. While final measurements are often classical in nature, it is possible that the space of probability distributions that are accessible by quantum evolution is distinguishable from probabilities defined by purely classical evolution. In fact, this is a statement of the concept of a quantum advantage in reverse; the gates in a computer implement the time evolution of the system and an outcome of that evolution, that would be impossible classically, is both a test of the quantum nature of the evolution and a quantum advantage.

In this paper, we argued that unitary quantum and classical Hamiltonian evolution cannot match both two point statistics and higher-point correlations simultaneously. If we hold the two-point statistics of a small number of commuting observables fixed, the classical theory will necessarily contain a number of states with frequencies that are not accessible as fluctuations from the quantum vacuum. A consequence is that these classical states can become degenerate, leading to additional classical correlations off-resonance and oscillations and eventual dephasing of the correlators on resonance. 

These results generalized prior work that showed classical correlators give rise to poles at physical momenta for cosmological correlators. These poles were shown to be precisely those associated with the breakdown of conventional perturbation theory and are resolved by using degenerate perturbation theory. In this sense, we find a direct connection between the quantum tests of the origin of structure in the universe, which relied on quantum field theory and relativistic causality, and more basic tests of quantum mechanics assuming only unitary local time evolution.

The ultimate goal of this program is to develop an understanding of quantum time-evolution at the same level of generality as Bell's inequalities. The observations made in terms of correlations still rely on the structure of Hamiltonian evolution that one does not directly observe. One may certainly wonder if there exists local stochastic models that can reproduce quantum vacuum correlations but do not have a local Hamiltonian description. This is no different than the challenge of finding a general theory of quantum advantage. Yet, with the development of tools like the cosmological bootstrap~\cite{Baumann:2022jpr}, the idea that time evolution can be inferred directly from the structure of correlations is potentially within reach. The possibility that such insights have deep connections well beyond cosmology suggests new directions and opportunities in our quest to understand the quantum nature of the Universe.

\paragraph{Acknowledgments}
We are grateful to Guillermo Ballesteros Martinez, Dan Carney, JJ Carrasco, Jesús Gambín, Peter Graham, Austin Joyce, Aurora Ireland, Jai Grewal, Tongyan Lin, Alejandro Perez Rodriguez, Rafael Porto, and Eva Silverstein for discussion. DG would particularly like to thank Dan Carney for raising the question of the connection between graviton detection and SQU proposal which inspired some of this project. DG and KG are supported by the US~Department of Energy under grant~\mbox{DE-SC0009919}. QZ was supported by a SPURS awards from UC San Diego.

\appendix


\section{Unitarity in Detail}\label{app:unitarity}

For quantum evolution, unitarity is imposed by requiring the Hamiltonian $H$ is hermitian. For classical systems, we want unitarity to enforce conservation of information. This is demanded by requiring the classical phase space flow generated by the Hamiltonian to be area conserving and time reversible. A necessary and sufficient condition for this is 
\begin{equation}
    \{x(t), p(t)\} = 1 \ ,
\end{equation}
where $\{a, b\}$ is the classical Poisson bracket, and $x, p$ are the generalized coordinates and momenta. The  important point in particular is that the Poisson bracket cannot be time-dependent. Time dependence of the Poisson bracket leads to change in phase space area over time, which implies that information is not conserved and the evolution is not invertible.

The time independence of the commutator constrains the form of $x(t), p(t)$ in the following way. Suppose we break up $x(t), p(t)$ into its respective Fourier modes. For simplicity, let us take that only two modes are present : 
\begin{equation}
    x(t) = a e^{i \omega_1 t} + b e^{i \omega_2 t} \qquad p(t) = m \dot{x} = im (a \omega_1 e^{i \omega_1 t}  + b \omega_2 e^{i \omega_2 t}) \ .
\end{equation}
Then, requiring $\{x(t), p(t)\} = 1$ enforces
\begin{equation}
    i\omega_2\{a, b\} e^{i (\omega_1 + \omega_2)t} + i\omega_1 \{b, a\} e^{i (\omega_1 + \omega_2)t} = 1 \ .
\end{equation}
For the Poisson bracket to be time independent, we see that require that $\omega_1 = - \omega_2$. Thus, we have that 
\begin{equation}
    x(t) = a e^{i \omega t} + b e^{-i \omega t} \ .
\end{equation}
In general, we must have that 
\begin{equation}
    x(t) = \sum_i (a_i e^{i \omega_i t} + b_i e^{-i \omega_i t}) \ .
\end{equation}
If we think of $e^{i \omega t}$ as ``lowering the energy", then this statement implies that classically, if we allow raising of energies, unitarity enforces that we also have to allow lowering of energies.

\section{Deriving the Resonant Hamiltonian}\label{app:resonant}

The resonant Hamiltonian is the classical equivalent of degenerate perturbation theory. The key idea is that given a list of frequencies that satisfy a resonant condition, we can perform a series of canonical transformations to remove the effect of time evolution that only shifts that individual frequencies but has no resonant effect of mixing the frequencies while conserving energy. The resonant Hamiltonian is the contribution that survives, but because it cannot be removed in this way, and because it leads to non-trivial time evolution that mixes all the frequencies.

\subsection{Illustrative Example}

Rather than present a general construction, we will work out the procedure in detail for the example in Section~\ref{subsec:spin}. We start with a Hamiltonian write in action angle variables, $I_i,\theta_i$ for $i=1,2,3$ as
\begin{align}
H =& \sum_{i=1}^3 \omega_i I_i + H_{\rm int}(I_i, \theta_i) \\
H_{\rm int}(I_i, \theta_i) &= g I_1 I_2 I_3 \cos(\theta_1) \cos(\theta_2) \cos(\theta_3) \\
=& \frac{g I_1 I_2 I_3}{4} \big(\cos(\theta_1+\theta_2+\theta_3) + \cos(\theta_1-\theta_2 -\theta_3) \\
&+\cos(\theta_1-\theta_2 +\theta_3)+\cos(\theta_1+\theta_2 -\theta_3) \big) \ .
\end{align}
Now we define a canonical transformation defined by a generating functional $F(\theta,I')$ so the Lagrangian is unchanged $I\dot \theta- H(\theta, I)=  I' \dot \theta' - H(\theta', I')$. By the usual type two construction
\beq
F = \theta I' + G(\theta, I') \ ,
\eeq
so that 
\beq
I =\frac{\partial}{\partial \theta } F  = I' +\frac{\partial}{\partial \theta } G \qquad \theta' =\frac{\partial}{\partial I' } F  =\theta +\frac{\partial}{\partial I' } G \ .
\eeq
With this definition, we see that Lagrangian, $\L$ changes by a total derivative 
\beq
\L(\theta,I)  -\L'(\theta',I') = I \dot\theta - I' \dot \theta' = \frac{d}{dt}\left( F(\theta, I') - \theta' I'\right) \ .
\eeq
Now we want to pick $G$ to eliminate the interactions that involve the high frequency, $\theta_1+\theta_2+\theta_3 \approx (\omega_1+\omega_2 +\omega_3)t$, so we can choose
\beq
G = - \frac{g}{4(\omega_1+\omega_2+\omega_3)} I'_1 I'_2 I'_3 \sin(\theta_1+\theta_2+\theta_3) \ .
\eeq
It is straightforward to check that under this change 
\beq
H_{\rm int}(I'_i,\theta_i') =\frac{g I'_1 I'_2 I'_3}{4} \big( \cos(\theta'_1-\theta'_2 -\theta'_3)
+\cos(\theta'_1-\theta'_2 +\theta'_3)+\cos(\theta'_1+\theta'_2 -\theta'_3) \big) +{\cal O}(g^2)\ .
\eeq
It is straightforward to show that we can remove any term involving $g C(I_i) \cos(k_1 \theta_1+k_2 \theta_2 + k_3 \theta_3)$ using the generating function with 
\beq\label{eq:canonical}
G = -\frac{g}{k_1\omega_1+k_2 \omega_2+k_3 \omega_3} C(I'_i) \sin(k_1 \theta_1+k_2 \theta_2+k_3 \theta_3) \ .
\eeq
This works as long as $g/(k_1\omega_1+k_2 \omega_2+k_3 \omega_3) \ll 1$, which clearly fails for the resonant interactions. As a result, by this sequence of canonical transformations, we see that only the resonant terms need to be solved explicitly.

Now suppose our observables are $X_i= I_i \cos\theta_i$. These operators will be shifted by this canonical transformation as
\beq
X_i = X'_i +\partial_{I_i} X \delta I_i  + \partial_{\theta_i} X_i \delta \theta_i = X'_i - \{ X_i, G \}  +{\cal O}(g^2) \ ,
\eeq
where $\delta I_i = \partial_{\theta_i} G$ and $\delta \theta_i= - \partial_{I'_i} G$. For this particular canonical transformation, we then find 
\beq
\delta X_i = X_i-X'_i = -\frac{g}{4 (\omega_1+\omega_2+\omega_3)} I_1 I_2 I_3 \left(\cos \theta_i \cos \theta_T +\sin \theta_i \sin \theta_T \right) \ ,
\eeq
where $\theta_T = \theta_1+\theta_2+\theta_3$. As a result, we see that 
\bea
\langle X_1 X_2 X_3 \rangle &=& \langle X'_1 X'_2 X'_3 \rangle\\
&&- \frac{g (I_1I_2 I_3)^2}{16(\omega_1+\omega_2+\omega_3)}\left( \sum_{i=1}^3 \frac{1}{I_i}(1+ \cos2 \theta_{i+1} +\cos 2 \theta_{i+2} + \cos(2(\theta_{i+1} +\theta_{i+2})) \right)  \ , \nonumber
\eea
where we defined the index mod(3) so that $i \equiv i+3$. We see we recover the total energy pole term from the canonical transformation. The time dependent terms will vanish as $t \to \infty$ due to dephasing.

Taking care of the all the non-resonant energy poles, we ultimately get that the shifted operators are given by 
\begin{equation}
 \begin{aligned} 
X_{1, \mathrm{old}} &= X_{1} - g \frac{I_{1}I_{2}I_{3}}{4} \left( \frac{cos(\theta_{2} + \theta_{3})}{\omega_{1} + \omega_{2} + \omega_{3}} + \frac{\cos(\theta_{2} - \theta_{3})}{\omega_{1} - \omega_{2} + \omega_{3}} + \frac{\cos(\theta_{2} - \theta_{3})}{\omega_{1} + \omega_{2} - \omega_{3}} \right)\\
X_{2, \mathrm{old}} &= X_{2} - \frac{gI_{1}I_{2}I_{3}}{4} \left(  \frac{\cos(\theta_{1} + \theta_{3})}{\omega_{1} + \omega_{2} + \omega_{3}} -\frac{\cos(\theta_{1} + \theta_{3})}{\omega_{1} - \omega_{2} + \omega_{3}} + \frac{\cos(\theta_{1} - \theta_{3})}{\omega_{1} + \omega_{2} - \omega_{3}} \right)\\
X_{3, \mathrm{old}} &= X_{3} - \frac{gI_{1}I_{2}I_{3}}{4}\left(  \frac{\cos(\theta_{1} + \theta_{2})}{\omega_{1}+ \omega_{2} + \omega_{3}} + \frac{\cos(\theta_{1} - \theta_{2})}{\omega_{1} - \omega_{2} + \omega_{3}} - \frac{\cos(\theta_{1} + \theta_{2})}{\omega_{1} + \omega_{2} - \omega_{3}} \right) \ ,
 \end{aligned} 
\end{equation}
which leads to 
\begin{equation}
 \begin{aligned} 
&\langle X_{1} X_{2} X_{3}\rangle  \to \langle X_{1} X_{2} X_{3} \rangle  - \\
&\frac{g(I_{1}^{2} I_{2}^{2} I_{3}^{2})}{16} \bigg(  \frac{1}{I_{1}}\left[  \frac{A^+_{23}}{D_{1}} + \frac{A^-_{23}}{D_{2}} + \frac{A^-_{23}}{D_{3}}  \right] + \frac{1}{I_{2}}\left[  \frac{A^+_{13}}{D_{1}} - \frac{A^+_{13}}{D_{2}} + \frac{A^-_{13}}{D_{3}}  \right] \\
& \qquad \qquad + \frac{1}{I_{3}}\left[  \frac{A^+_{12}}{D_{1}} + \frac{A^-_{12}}{D_{2}} - \frac{A^+_{12}}{D_{3}} \right]  \bigg) \ ,
 \end{aligned} 
\end{equation}
where we have defined
\begin{equation}
    D_{1} = \omega_{1} + \omega_{2} + \omega_{3} \qquad D_{2} = \omega_{1} - \omega_{2} + \omega_{3} \qquad D_{3} = \omega_{1} + \omega_{2} - \omega_{3} \ ,
\end{equation}
and 
\begin{equation}
    A_{ij}^\pm = 1 + \cos 2 \theta_{i} + \cos 2 \theta_{j} + \cos(2 \theta_{i} \pm 2 \theta_{j}) \ .
\end{equation}
If we take cosine terms to dephase, then we get 
\begin{equation}
 \begin{aligned} 
-g \frac{(I_{1}I_{2}I_{3})^{2}}{16} \left[  \frac{\frac{1}{I_{1}} + \frac{1}{I_{2}} + \frac{1}{I_{3}}}{\omega_{1} + \omega_{2} + \omega_{3}} + \frac{\frac{1}{I_{1}} - \frac{1}{I_{2}} + \frac{1}{I_{3}}}{\omega_{1} - \omega_{2} + \omega_{3}} + \frac{\frac{1}{I_{1}} + \frac{1}{I_{2}} - \frac{1}{I_{3}}}{\omega_{1} + \omega_{2} - \omega_{3}} \right] 
 \end{aligned} 
\end{equation}
as our matching condition.

Now we can consider what happens to the resonant interaction as we approach the pole. After all the canonical transformation described by Equation~(\ref{eq:canonical}), we will be left with an action
\beq
H= \omega_1 I_1' + \omega_2 I'_2 + \omega_3 I'_3 + \frac{g}{4}I_1' I_2' I_3'\cos \gamma \ ,
\eeq
where $\gamma = \theta_1-\theta_2 -\theta_3$. We can change variables to $\gamma$, $J = I_1'$, $P_2=I_1'+I_2'$ and $P_3= I_1'+I_3'$ so that 
\beq
H = \Delta J + \omega_2 P_2 +\omega_3 P_3 + \frac{g}{4} J(P_2-J)(P_3-J)\cos\gamma \ .
\eeq
Here $P_2$ and $P_3$ are constants under that equations of motion. Furthermore, $H$ is itself conserved by the equation. Using these conserved quantities, we can explicitly solve for the correlator flowing under the resonant Hamiltonian. Thus, in terms of the original correlator, we have that 
\begin{equation}
    \langle X_1 X_2 X_3 \rangle = \langle X_1 X_2 X_3 \rangle_{\rm matching} + \frac{1}{4} \langle I_1' I_2' I_3' \cos \gamma \rangle \ ,
\end{equation}
and so in the resonance Hamiltonian picture, we are left with $I_1' I_2' I_3' \cos \gamma$. Then, taking the initial condition $\langle I_1' I_2' I_3' \cos \gamma \rangle(t= 0) = 0$ we have that 
\begin{equation}
    I_1' I_2' I_3' \cos \gamma = \frac{4}{g} (H - \omega_2 P_2 - \omega_3 P_3- \Delta J) = -\frac{4 \Delta}{g} (J(t) - J(0)) \ .
\end{equation}
We see that this is a non-perturbative solution, which vanishes on resonance, $\Delta \to 0$. At exact resonance, we have that $H - \omega_2 P_2 - \omega_3 P_3= I'_1 I_2' I_3' \cos \gamma$ is conserved, and hence we get $\langle I_1' I_2' I_3' \cos \gamma \rangle (t) = \langle I_1'I_2'I_3' \cos \gamma \rangle (0)  = 0$. For finite but small $\Delta$, we can solve this equation exactly. We have that 
\begin{equation}
\begin{aligned}
    \dot{J} &= \frac{g}{4} J (P_{2} - J)(P_{3} - J) \sin \gamma \\ \dot{\gamma} &= \Delta + \frac{g}{4} ((P_{2} - J)(P_{3} - J) - J(P_{3} - J) - J(P_{2} - J))\cos \gamma \ ,
\end{aligned}
\end{equation}
where $P_2$, $P_3$ are constants. Then, for given constant $P_2, P_3, H$ values, we can solve for $J(t), \gamma(t)$ exactly. We have that calling 
$C = H - \omega_2 P_2 - \omega_3 P_3$, we get that 
\begin{equation}
    \dot{J} =  \pm\sqrt{ \frac{g^{2}}{16} J^{2} (P_{2} - J)^{2} (P_{3} - J)^{2} - (C - \Delta J)^{2} }  \ ,
\end{equation}
and so we have that 
\begin{equation}
    2\int^{J_+}_{J_-} \frac{dJ}{\sqrt{ \frac{g^{2}}{16} J^{2} (P_{2} - J)^{2} (P_{3} - J)^{2} - (C - \Delta J)^{2} }} = T \ .
\end{equation}
Evaluating this at the turning points tells us that the value of $J$ oscillates with a frequency given by $\omega = 2\pi/ T$. As we can see, the frequency itself is dependent on the amplitude turning points, and hence generically the frequency is amplitude dependent. Thus, we have that from the usual dephasing argument, we get
\begin{equation}
    \langle J(t)  \rangle \to J_* \ ,
\end{equation}
which is the stationary value. Thus, we see that slightly off-resonance we have 
\begin{equation}
    \langle I_1' I_2' I_3' \cos \gamma \rangle \to -\frac{4 \Delta}{g} J_*
\end{equation}
settles to some stationary value, which at exact resonance goes to 0.

\subsection{Generalities}\label{subapp:generalities}

The general construction of the resonant Hamiltonian can be understood as follows: given $H_0=\sum_i \omega_i I_i$ and angle dependent interaction terms $H_{\rm int} = g h (\theta_j, I_j)$, for some function $h$, we can use canonical transformations to remove the high frequency components order by order in the coupling $g$. Intuitively, this is the idea that we can decompose $h$ as a fourier series in $\theta_j$, e.g.~for $j=1,2,3$ we can write
\beq
h(\theta_j, I_j) = h_0(I_j) + \sum_{n_1,n_2,n_3\neq 0} e^{i n_1 \theta_1 + i n_2 \theta_2 +i n_3 \theta_3}h_{n_1,n_2,n_3}(I_i) \ .
\eeq
Using the canonical transformation described above, we can clearly remove terms when $\sum_j n_j \omega_j \neq 0$. Intuitively, this is similar to the expectation that we can average in time to eliminate the high frequencies. The canonical transformation achieves the same goals without the need for some ad hoc averaging procedure.

Two kinds of terms remain after this transformation. We have focused on the resonant terms where $\sum_i n_i \omega_i =0$ because they introduce non-trivial mixing between the modes. In short, at ${\cal O}(g)$, the resonant Hamiltonian is simply
\beq
H_{\rm res}= g h_{\rm res}= g \sum_{n_1,n_2,n_3 \neq 0}h_{n_1,n_2,n_3}(I'_j) e^{i \sum_j n_j \theta_j} \delta_{\sum_j n_j \omega_j , 0} \ .
\eeq
This is the essential result that gives rise to the time dependence of the correlator and ultimate the dephasing over time.

In the above example $h_0(I_j) = 0$. As a result, we did not see any since to the frequencies $\omega_j$. This is not true in general, and when $h_0(I_j) \neq 0$, the evolution of the angle variables shifts at ${\cal O}(g)$,
\beq
\dot \theta_j = \omega_j + g\frac{\partial}{\partial I'_j} (h_0(I'_j) + h_{\rm res}(I',\theta_j) \ .
\eeq
As a result, we see that even if $h_{\rm res}= 0$, $h_0 \neq 0$ is sufficien to generate a $I_j$-dependent shift to the frequencies
\beq
\omega_j' \approx \omega_j + g\frac{\partial}{\partial I'_j} h_0(I'_j) \ .
\eeq
Since $I'_j$ is a random variable, this shift also leads to the decay of correlations after averaging over the amplitudes. 

This feature of the classical theory is also not reproduced by the ground state of a quantum system. Given a non-degenerate system and an interaction Hamiltonian $H_{\rm int}$. For illustration, we can take a single harmonic oscillator with an interaction term
\beq
H = \omega a^\dagger a + g (a^\dagger a)^m \ .
\eeq
With $g=0$ the states are labeled by an integer $n$ with energies $E_n = n \omega$ (after shifting the energy of the ground state to zero, for convenience). 

In the quantum description, the states and energy levels are shifted so that 
\beq
E_n = \omega n+ g n^m \ .
\eeq
Taking the classical limit, $\omega \to 0$ and $n  \to \infty$, with $I = \kappa n$ and $\omega' = \omega / \kappa$ as our classical variables so that
\beq
H \to \omega' I + g \kappa^m I^m \ .
\eeq
In this description, treating $I$ as a random variable requires that our theory is in a mixed state
\beq
\rho = \sum_n p_n |n\rangle \langle n | \ .
\eeq
In the quantum description, the energies / frequencies are shifted by a constant amount. The origin of the nonlinear relationship between frequency and amplitude of the classical theory is that a varying $I$ at fixed $\omega'$ corresponds to changing $n$ and $\kappa$, which doesn't hold the energy of the quantum theory fixed.

\section{Classical vs Quantum on resonance} \label{app:CM_vs_QM_res_corr}

We have seen that the resonance poles can be used as a way to probe the difference between the classical and the quantum theory. However, when we go to resonance, the pole structure vanishes and needs to be resummed. In this section, we want to explain, in the two examples of spins and harmonic oscillators that we have seen in \ref{subsec:spin} and \ref{subsec:harmonic}, how the resonant Hamiltonian leads to differences between the classical and quantum theories in a way that is observable by measuring the correlators of the theory. 

\subsection{Harmonic Oscillators}

Consider the case of three coupled oscillators as in \ref{eq:Hp3}
\begin{equation}
    H_{\rm int} = g x_1 x_2 x_3 \ ,
\end{equation}
and impose the $\omega_1 = \omega_2 + \omega_3$ resonance condition. The free theory is given by 
\begin{equation}
    x = \sigma (a e^{-i \omega t} + a^\dagger e^{i \omega t}) \qquad \sigma = \frac{1}{\sqrt{2 \omega_i}} 
\end{equation}
for the quantum case, where $a, a^\dagger$ are operators, with $[a, a^\dagger] = 1, \langle a a^\dagger \rangle = 1$, while for the classical case we have
\begin{equation}
    x = \sigma (a e^{-i \omega t} + a^* e^{i \omega t}) \ ,
\end{equation}
where $a, a^*$ are random variables sampled from a distribution, with $\langle a a^* \rangle = 1/2$. For the case of resonance, if we are sitting on the pole, we generically have classically that the amplitude becomes time dependent. Denoting this case with $A(t)$, we have that on resonance, we get
\begin{equation}
    x = \sigma (A e^{-i \omega t} + A^* e^{i \omega t}) \ .
\end{equation}
Then, the resonant Hamiltonian is 
\begin{equation}
    H_{\rm res} = \tilde{g} (A_1 A_2^* A_3^* + A_1^* A_2 A_3) \qquad \tilde{g} = g \sigma_1 \sigma_2 \sigma_3 \ .
\end{equation}
The equations of motion is given by 
\begin{equation}
    i \dot{A}_1 = \tilde{g} A_2 A_3 \qquad i \dot{A}_2 = \tilde{g} A_1 A_3^* \qquad i \dot{A}_3 = \tilde{g} A_1 A_2^* \ .
\end{equation}
This is the resonant Hamiltonian picture, and we will use this to build concrete observables which will test the classical versus quantum correlators. 

\subsection{Building an observable}

For the above Hamiltonian, define $I_i = |A_i|^2$. And define ${\cal J} = \rm{Im}(A_1^* A_2 A_3)$. Then, we have that 
\begin{equation}
    \dot{I}_1 = 2 \tilde{g} {\cal J} \qquad \dot{I}_2 = - 2 \tilde{g} {\cal J} \qquad \dot{I}_3 = -2 \tilde{g} {\cal J} \ .
\end{equation}
The conserved combinations are
\begin{equation}
    I_1 + I_2 \qquad I_1 + I_3 \ .
\end{equation}
Now, we see that 
\begin{equation}
    \dot{\cal J} = \kappa K \qquad K \equiv I_2 I_3 - I_1 I_2 - I_1 I_3 \ .
\end{equation}
We will use this observable to distinguish classical and quantum signatures. In the long time limit, we expect classically that 
\begin{equation}
    \langle \dot{\cal J} \rangle_{\rm cl} = 0 \ .
\end{equation}
This is because the late time limit of $\langle \cal J \rangle$ tends to a constant from dephasing, $\langle {\cal J} \rangle_{\rm cl} \to \cal J_* $, and hence $\langle{\cal \dot{J}} \rangle$ vanishes . This implies that classically 
\begin{equation}
    \langle I_2 I_3 - I_1 I_2 - I_1 I_3 \rangle_{\rm CM} \to 0 \ .
\end{equation}
We will now show that quantum mechanically, 
\begin{equation}
    \langle I_2 I_3 - I_1 I_2 - I_1 I_3 \rangle_{\rm QM} \to -\frac{1}{4} \ .
\end{equation}
Remembering that, $I_i = |A|_i^2$, we have $x^2 = 2 \sigma^2 I_i$. Using this relation, we have that 
\begin{equation}
     \langle I_2 I_3 - I_1 I_2 - I_1 I_3 \rangle = \left\langle  \frac{x_{2}^{2}}{2 \sigma_{2}^{2}} \frac{x_{3}^{2}}{2 \sigma_{3} ^{2}} - \frac{x_{1}^{2}}{2\sigma_{1}^{2}} \frac{x_{2}^{2}}{2 \sigma_{2}^{2}} - \frac{x_{1}^{2}}{2 \sigma_{1} ^{2}} \frac{x_{3}^{2}}{2 \sigma_{3}^{2}} \right\rangle \ .
\end{equation}
As we have showed already, this value is 0 classically. When we substitute $I \to x^2$, we are neglecting the corrections that fast modes produce. At best, they produce a ${\cal O}(g)$ correction. Thus, we have 
\begin{equation}
    \left\langle  \frac{x_{2}^{2}}{2 \sigma_{2}^{2}} \frac{x_{3}^{2}}{2 \sigma_{3} ^{2}} - \frac{x_{1}^{2}}{2\sigma_{1}^{2}} \frac{x_{2}^{2}}{2 \sigma_{2}^{2}} - \frac{x_{1}^{2}}{2 \sigma_{1} ^{2}} \frac{x_{3}^{2}}{2 \sigma_{3}^{2}} \right\rangle_{\rm CM} = 0 + {\cal O}(g) \ .
\end{equation}
On the other hand the quantum vacuum has no resonance. Thus, the quantum vacuum is the same as the free Gaussian theory upto ${\cal O}(g)$ corrections. For the harmonic case, since $\langle x^2\rangle = \sigma^2$, we get that 
\begin{equation}
    \left\langle  \frac{x_{2}^{2}}{2 \sigma_{2}^{2}} \frac{x_{3}^{2}}{2 \sigma_{3} ^{2}} - \frac{x_{1}^{2}}{2\sigma_{1}^{2}} \frac{x_{2}^{2}}{2 \sigma_{2}^{2}} - \frac{x_{1}^{2}}{2 \sigma_{1} ^{2}} \frac{x_{3}^{2}}{2 \sigma_{3}^{2}} \right\rangle_{\rm QM} = -\frac{1}{4} + {\cal O}(g) \ .
\end{equation}
So, we see that the resonance Hamiltonian produces an ${\cal O}(1)$ difference in relations between correlators. While we built an observable which seems special to the three oscillator case with $1 \to 2$ resonance, we can build similar correlators for any $m \to n$ resonance case for harmonic oscillators. Thus, this construction gives a blueprint as to how to find exact correlators which can help differentiate the classical distribution from the quantum vacuum. Moreover, it shows that the existence of resonance separates out the space of probability distributions of the classical theory from the quantum theory, atleast in these simple solvable models.

\subsection{Spin Example}

Now, we want to return to our example on spin states to build a similar observable. We have that the resonant Hamiltonian is given by
\begin{equation}
    H = \omega_{1} I_{1} + \omega_{2} I_{2} + \omega_{3} I_{3} + \frac{g}{4} I_{1}I_{2}I_{3} \cos \gamma \ .
\end{equation}
We now want to work at exact resonance. There we have that 
\begin{equation}
    \dot{I}_{1} = -\frac{g}{4} I_{1}I_{2}I_{3} \sin \gamma \qquad \dot{I}_{2} = \frac{g}{4} I_{1}I_{2}I_{3} \sin \gamma \qquad \dot{I}_{3} = \frac{g}{4} I_{1}I_{2}I_{3} \sin \gamma \ .
\end{equation}
Denoting ${\cal J} = I_1 I_2 I_3 \sin \gamma$, we have 
\begin{equation}
    \dot{\cal J} = \frac{g}{4}(I_1 I_2^2 I_3^2 - I_1^2 I_2 I_3^2 - I_1^2 I_2^2 I_3) \ .
\end{equation}
Again, we expect that in the classical ensemble, this will generically dephase. Now, since $I_i$ here are just the magnitude of the spin variables, and since $S^2 = I^2 \cos^2 \theta \to I^2/2$ we have that the appropriate observable is
\begin{equation}
    \langle\dot{\cal J} \rangle \to \bigg\langle g( |S_1| S_2^2 S_3^2 - S_1^2 |S_2| S_3^2 - S_1^2 S_2^2 |S_3|) \bigg\rangle \ .
\end{equation}
We always have that the magnitude of the spin variable will be $S_i = 1$. Hence, by dephasing arguments similar to the harmonic case, we expect that classically
\begin{equation}
    \langle S_2^2 S_3^2 - S_1^2 S_3^2 - S_1^2 S_2^2\rangle_{\rm CM} \to 0 + {\cal O}(g)\ ,
\end{equation}
in the late time limit, with the ${\cal O}(g)$ corrections coming from the non-resonant fast modes. However, quantum mechanically, we will have that the ground state to leading order is the unperturbed ground state, and satisfies
\begin{equation}
    \langle S_2^2 S_3^2 - S_1^2 S_3^2 - S_1^2 S_2^2\rangle_{\rm QM} \to -\frac{1}{4} + {\cal O}(g)\ .
\end{equation}
Hence, we see that similar to the harmonic oscillator case, the classical distribution and quantum vacuum can be differentiated via measuring these correlators.

\addcontentsline{toc}{section}{References}
\small
\bibliographystyle{utphys}
\bibliography{Refs}

\end{document}